\documentclass[aps,pra,twocolumn,superscriptaddress,longbibliography]{revtex4-2}
\usepackage{amsfonts,amsmath,amssymb,epsfig}
\usepackage{epsfig}
\usepackage{graphicx}
\usepackage[dvipsnames]{xcolor}
\usepackage{color}
\usepackage{soul}

\usepackage{epstopdf}
\usepackage{array}
\usepackage{tabularx}
\usepackage[utf8]{inputenc}
\usepackage[T1]{fontenc}
\usepackage[colorlinks,linkcolor=blue, citecolor=blue,
urlcolor=blue]{hyperref}
\newcommand\eref[1]{(\ref{#1})}

\newcommand\fref[1]{Fig.~(\ref{#1})}

\begin{document}
\title{Thermal Decay of Planar Jones-Roberts Solitons}
\author{Nils A. Krause}
\affiliation{Department of Physics, University of Otago, Dunedin, New Zealand}
\affiliation{Dodd-Walls Centre for Photonic and Quantum Technologies}
\author{Ashton S. Bradley}
\affiliation{Department of Physics, University of Otago, Dunedin, New Zealand}
\affiliation{Dodd-Walls Centre for Photonic and Quantum Technologies}

\date{\today}
\begin{abstract}
Homogeneous planar superfluids exhibit a range of low-energy excitations that also appear in highly excited states like superfluid turbulence. In dilute gas Bose-Einstein condensates, the Jones-Roberts soliton family includes vortex dipoles and rarefaction pulses in the low and high velocity regimes, respectively. These excitations carry both energy and linear momentum, making their decay characteristics crucial for understanding superfluid dynamics. In this work, we develop the theory of planar soliton decay due to thermal effects, as described by the stochastic projected Gross-Pitaevskii theory of reservoir interactions. We analyze two distinct damping terms involving transfer between the condensate and the non-condensate reservoir: particle transfer that also involves energy and usually drives condensate growth, and number-conserving energy transfer. We provide analytical treatments for both the low and high velocity regimes and identify conditions under which either mechanism dominates. Our findings indicate that energy damping prevails at high phase space density. These theoretical results are supported by numerical studies covering the entire velocity range from vortex dipole to rarefaction pulse. We use interaction energy to characterize rarefaction pulses, analogous to the distance between vortices in vortex dipoles, offering an experimentally accessible test for finite temperature theory in Bose-Einstein condensates.
 \end{abstract}
\flushbottom
\maketitle
\section{Introduction}
Homogeneous superfluid Bose-Einstein condensates (BECs) support a diverse range of excitations~\cite{svistunov_superfluid_2015} from weakly interacting Bogoliubov phonons to quantum vortices, dark solitons~\cite{pitaevskii_bose-einstein_2003}, and Jones-Roberts solitons (JRS)~\cite{jones_motions_1982}. JRS range from vortex rings or dipoles at low velocities~\cite{pismen_vortices_1999} to rarefaction\hl{} pulses approaching the speed of sound~\cite{tsuchiya_solitons_2008}. These solitons form during the decay of high energy BEC dynamics, playing a crucial role in superfluid relaxation and energy transport. Recent planar turbulence experiments~\cite{neely_characteristics_2013,gauthier_giant_2019,johnstone_evolution_2019,galka_emergence_2022,gazo_universal_2023,karailiev_observation_2024} have underscored the need for a comprehensive understanding of the interactions and decay mechanisms of JRS.  

As two counter-rotating quantum vortices in a two-dimensional BEC approach each other closely they annihilate~\cite{mironov_propagation_2012,rorai_propagating_2013,kwon_relaxation_2014,kwon_sound_2021}. However, instead of immediately disintegrating into sound waves they form a compressible solitary wave, the so-called rarefaction pulse. These two excitations were given a unified description as nonlinear eigenstates of the Gross-Piteaveskii equation in a moving coordinate frame by Jones and Roberts~\cite{jones_motions_1982}. Despite the success of zero temperature theory in describing JRS, their dissipation mechanisms have received little attention. Experimental evidence supports steady motion on typical timescales of BEC dynamics \cite{meyer_observation_2017}, consistent with theoretical predictions of stability at very low temperatures \cite{kuznetsov_two-_1982,tsuchiya_stability_2007,tsuchiya_solitons_2008}. As there is no inherent instability (at least as long as the third dimension is sufficiently suppressed \cite{kuznetsov_two-_1982,kuznetsov_instability_1995}), thermal effects are the primary mechanism of decay.

The truncated Wigner phase space method provides a practical means to studying dissipative processes in BEC. It is well suited to making semi-classical approximations, while describing the influence of dissipation and noise on BEC excitations\cite{blakie_dynamics_2008}. Recent progress on the theory of vortex damping using the stochastic projected Gross-Pitaevskii equation (SPGPE) \cite{gardiner_stochastic_2003,bradley_bose-einstein_2008,rooney_stochastic_2012} gives clear indication that dissipation is driven by energy exchange with the non-condensate~\cite{mehdi_mutual_2023}, in agreement with experiments~\cite{kwon_relaxation_2014,moon_thermal_2015,kwon_sound_2021} and ZNG simulations \cite{jackson_finite-temperature_2009}; the energy-damping mechanism involves no particle exchange with the reservoir, in contrast to commonly used treatments of damping in Gross-Piteavskii theory~\cite{choi_phenomenological_1998}. In this work we extend the description to a treatment of JRS decay close to and after the critical velocity for dipole annihilation. Our analysis makes use of formal properties of JRS \cite{jones_motions_1982,kuznetsov_instability_1995} together with renormalised soliton Lagrangian perturbation theory~\cite{kivshar_lagrangian_1995} to find analytical expressions for their decay. Our treatment provides a clear comparison of the strength of two distinct decay mechanisms appearing in finite-temperature BEC theory. Significantly, we find that both rarefaction pulse and dipole decay are dominated by the energy damping process once the phase space density is significant. The decay rates predicted by energy damping are typically an order of magnitude faster than the predictions of number damping, consistent with recent theory of vortex energy damping~\cite{mehdi_mutual_2023}.

This paper is structured as follows. In section \ref{bkg} we collect essential background material for the theory of JRS and BEC dissipation. In section \ref{weakdamping} we derive analytical expressions for number and energy damping mechanisms, find conditions for each mechanism to dominate the dissipation of JRS, and estimate the JRS lifetime. In section \ref{numver} we present numerical simulations of the decay that are in good agreement with our analytical predictions. The interaction energy is emphasized as a good measure of the rarefaction pulse decay; we also use  use Pad{\'e} approximants to describe JRS damping close to dipole annihilation. Our conclusions are presented in section \ref{conc}.

\section{Background}\label{bkg}
In this section we collect a number of useful properties of JRS and reservoir theory that will be needed to treat their damping. 
\subsection{Jones-Roberts Solitons}
We consider a planar geometry, in which JRS are solutions of the zero temperature Gross-Pitaevskii equation (GPE) \cite{ginzburg_theory_1958-1,gross_hydrodynamics_1963}
\begin{align}
	\label{GPE}
	i\hbar\partial_t\psi(\textbf{r})=(\mathcal{L}_\text{GP}-\mu)\psi(\textbf{r}),
\end{align}
where $\psi$ is the condensate order parameter with normalisation $\int d^2\mathbf{r}|\psi(\mathbf{r})|^2=N$, $\mu$ is the chemical potential, and $\mathcal{L}_\text{GP}$ is the Gross-Pitaevskii operator
\begin{align}
	\mathcal{L}_\text{GP}\psi(\textbf{r})\equiv\left(-\frac{\hbar^2}{2m}\Delta+g|\psi(\textbf{r})|^2\right)\psi(\textbf{r}).
\end{align}
In two dimensions the cold-collision interaction parameter $g$ takes the form \cite{bradley_low-dimensional_2015}
\begin{align}
	g&=\frac{\sqrt{8\pi}\hbar^2a_\text{s}}{ml_z},
\end{align}
with $a_\text{s}$ the scattering length and $l_z$ the thickness of the atomic cloud in the $z$-direction. JRS are assumed to inhabit a uniform 2D superfulid with background density $n_0=\mu/g$. Choosing a coordinate system in which the solitons travel along the $x$-axis, JRS are defined as translating eigenstates of the GP-operator of the form
\begin{align}
	\label{solitonic}
	\psi(x,y,t)=\psi(x-x_\text{s}(t),y,0),
\end{align}
with $x_\text{s}$ the position of the soliton. The time and space variations are then related by
\begin{align}
	\label{soliton}
	\partial_t\psi(\textbf{r},t)=-v\partial_x\psi(\textbf{r},t).
\end{align}
We will assume the ideal soliton wavefunction $\psi$ to satisfy (\ref{solitonic}) and hence (\ref{soliton}) throughout this work. Dissipation mechanisms cause weak decay of the soliton, associated with slowly increasing $v(t)$ that can be treated as a weak perturbation of the ideal soliton behaviour.

\begin{figure}
	\centering
	\includegraphics[width=1\linewidth]{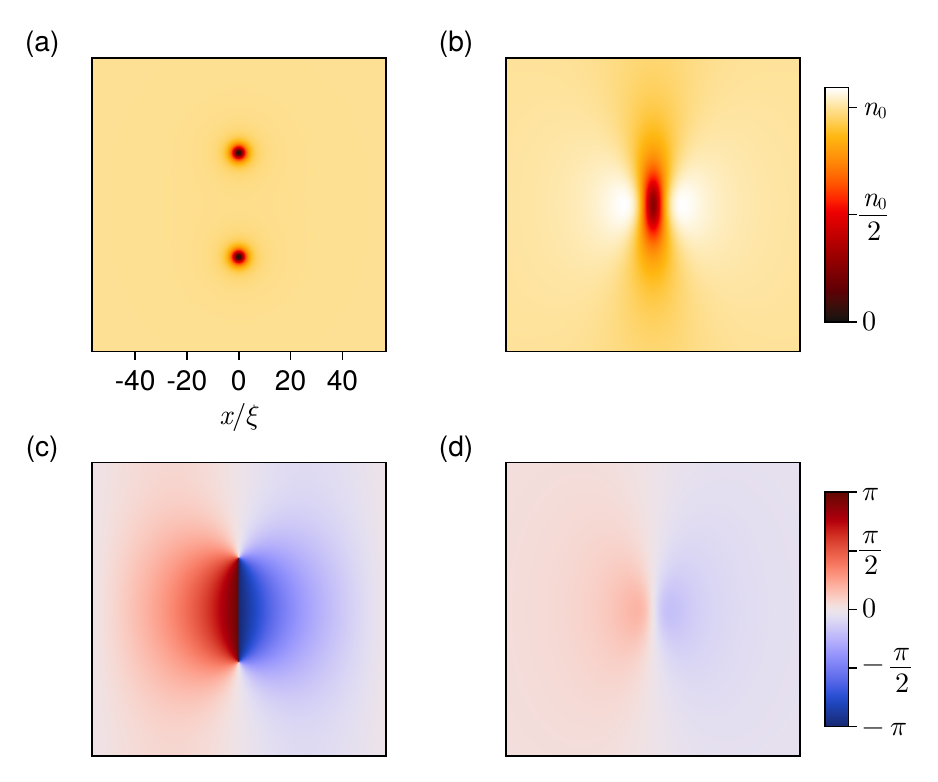}
	\caption{Density and phase plot of a planar Jones-Roberts soliton with a velocity below (\textbf{(a)} and \textbf{(c)}, respectively; the velocity is $v=0.05c$) and above (\textbf{(b)} and \textbf{(d)}, respectively; the velocity is $v=0.85c$) the critical velocity $v_\text{c}\simeq0.636c$ \cite{jones_motions_1986}. Below $v_\text{c}$ it consists of two counter-rotating vortices, with a $2\pi$-winding around each of them. Above, there remains only one minimum that does not reach to zero anymore. Additionally, the phase does not take all values between $0$ and $2\pi$. With further increasing velocity the JRS flattens and widens.}
	\label{fig:JRS}
\end{figure}
JRS arise for $v$ smaller than the speed of sound $c=\sqrt{gn_0/m}=\hbar/(\sqrt{2}m\xi)$. Here $\xi=\hbar/\sqrt{2m\mu}$ denotes the healing length. The family of JRS can be divided into two distinct regimes~\cite{jones_motions_1986}, as shown in Fig.~\ref{fig:JRS}. Below the critical velocity for vortex-antivortex annihilation, $v_\text{c}\simeq0.636c$, they represent a vortex dipole, consisting of two minima of zero density with a phase winding of $\pm 2\pi$ around each vortex. As $v$ approaches $v_\text{c}$ from below these vortices approach each other closely, leading to a depletion in the region between them and a breakdown of topological stability. For $v>v_\text{c}$ a single density minimum remains that no longer reaches zero. With increasing velocity, this rarefaction pulse grows larger in spatial extent and shallower, eventually vanishing as $v\to c$.
\subsection{Renormalised soliton integrals}
We will make use of a Lagrangian method to derive decay rates of the JRS. The Lagrangian in a homogenous Bose-Einstein condensate can be written as 
\begin{align}
	\begin{split}
		L=\int d^2\textbf{r}\bigg[&\frac{i\hbar}{2}(\psi^*\partial_t\psi-\psi\partial_t\psi^*)\left(1-\frac{n_0}{|\psi|^2}\right)\\
		&-\frac{\hbar^2}{2m}|\nabla\psi|^2-\frac{g}{2}(|\psi|^2-n_0)^2\bigg].
	\end{split}
\end{align}
where the infinite homogeneous background contribution has been subtracted~\cite{kivshar_lagrangian_1995}. For $\psi$ fulfilling (\ref{solitonic}) we can write the Euler-Lagrange equation (ELE)
\begin{align}
	\label{ELE}
	\frac{\partial L}{\partial x_\text{s}}-\frac{d}{dt}\frac{\partial L}{\partial \dot{x}_\text{s}}=0.
\end{align}
Since $L$ does not depend on the position of the JRS, we obtain for the left hand side of (\ref{ELE}) the negative time derivative of the momentum, $-\dot{P}$, with
\begin{align}
	P\equiv\frac{\partial L}{\partial \dot{x}_\text{s}}=\frac{i\hbar}{2}\int d^2\textbf{r}(\psi\partial_x\psi^*-\psi^*\partial_x\psi)\left(1-\frac{n_0}{|\psi|^2}\right).
\end{align}
Thus, the momentum of a JRS is conserved during its evolution. However, if there is damping, the momentum will diminish. In characterising the decay we will refer to the damping rate of the momentum in most of our results.

In deriving expressions for the damping rates, we will use a few integral relations, valid for quasi-solitary solutions of the GPE. To state them, let us define the total energy
\begin{align}
	\label{Etot}
	E_\text{tot}&=\int d^2\textbf{r}\bigg[\frac{\hbar^2}{2m}|\nabla\psi|^2+\frac{g}{2}(|\psi|^2-n_0)^2\bigg],
\end{align}
and interaction energy
\begin{align}
	E_\text{int}=\frac{g}{2}\int d^2\textbf{r}(|\psi|^2-n_0)^2.
\end{align}
Combining (\ref{soliton}) and (\ref{GPE}) gives
\begin{align}
	\label{stationary}
	-i\hbar v\partial_x\psi=(\mathcal{L}_\text{GP}-\mu)\psi.
\end{align}
If we multiply this equation by $x\partial_x\psi^*$, take the real part, integrate and perform integration by parts, we obtain \cite{jones_motions_1986}\footnote{Note that to derive this result we do not need to restrict ourselves to the two dimensional case. Hence, it also holds for solitons in one dimension and JRS in three dimensions.} 
\begin{align}
	\label{Integral-relation1}
	E_\text{tot}=\frac{\hbar^2}{m}\int d^2\textbf{r}|\partial_x\psi|^2.
\end{align}

If we instead multiply by $y\partial_y\psi^*$ and take again the real part, and integrate by parts, with the help of (\ref{Integral-relation1}) we now obtain the 2D result~\cite{jones_motions_1986}
\begin{align}
	\label{Integral-relation2}
	P=2\frac{E_\text{int}}{v}.
\end{align}
The momentum is hence proportional to the interaction energy in the high-velocity limit (in which $v\to c$ is approximately constant) and antiproportional to the velocity in the low-velocity limit (in which $E_\text{int}$ is constant for a vortex dipole). As we derive damping rates with regard to the momentum, this result will be of great use: instead of the more complicated and hardly measurable momentum we can characterise the decay by the easily accessible interaction energy and distance between the vortices respectively.

\subsection{Vortex dipole}
For widely separated vortices (low-velocity limit), their speed is given by the velocity field generated by the respective other vortex. This leads to a dependence on the distance of the vortices $d_\text{v}$ and the velocity of the dipole $v$ \cite{fetter_vortices_1966,lucas_sound-induced_2014}
\begin{align}
	\label{vd}
	v=\frac{\hbar}{md_\text{v}}.
\end{align}
In the dipole regime the total energy can be very well approximated as
\begin{align}
	\label{Etotd}
	E_\text{tot}=2\pi \frac{n_0\hbar^2}{m}\ln\left(\alpha\frac{d_\text{v}}{\xi}\right),
\end{align}
where the factor $\alpha$ was numerically found to be $\alpha=1.46$ \cite{cawte_snells_2019}~\footnote{Note the definition of healing length $\xi$ used in that work differs by a factor of $\sqrt{2}$.}. While the logarithmic scaling stems from the hydrodynamic limit that neglects the vortex core, the constant correction given by $\alpha$ accounts for the interaction and quantum pressure energies of each vortex core and can not be neglected for experimentally achievable dipole sizes (tens of healing length). The interaction energy can be approximated by  \cite{kuznetsov_instability_1995}
\begin{align}
	\label{Eintd}
	E_\text{int}\simeq\pi \frac{n_0\hbar^2}{m},
\end{align}
a constant independent of $d_{\mathrm v}$.
\subsection{Rarefaction pulse}
After the annihilation of the vortex dipole ($v=v_c$) the planar Jones-Roberts soliton moves on as a localised density dip, growing wider and shallower with increasing velocity. In the high-velocity limit ($v\lesssim c$) a useful quantity turns out to be the reciprocal Lorentz factor
\begin{align}
	\epsilon=\sqrt{1-\left(\frac{v}{c}\right)^2},
\end{align}
a small parameter as $v\to c$, allowing an expansion of the wavefunction in powers of $\epsilon$.  At leading order an analytical form of the JRS is known \cite{tsuchiya_solitons_2008}: it has density ($n(\textbf{r})=|\psi(\textbf{r})|^2$)
\begin{align}
	\label{rhohighv}
	n(\textbf{r})=n_0\left(1-4\epsilon^2\frac{3/2+\epsilon^4y^2/\xi^2-\epsilon^2x^2/\xi^2}{\left[3/2+\epsilon^4y^2/\xi^2+\epsilon^2x^2/\xi^2\right]^2}\right)+\mathcal{O}(\epsilon^3)
\end{align}
and phase
\begin{align}
	\label{theta}
	\theta(\textbf{r})=-2\sqrt{2}\epsilon\frac{\epsilon x/\xi}{3/2+\epsilon^4y^2/\xi^2+\epsilon^2x^2/\xi^2}+\mathcal{O}(\epsilon^2).
\end{align}
We will make use of this solution to analyze the damping behaviour in the high-velocity limit.

Direct calculation employing (\ref{rhohighv}) gives the interaction energy of a JRS at leading order
\begin{align}
	\label{Eintlinear}
	E_\text{int}=\frac{4\pi}{3}\frac{n_0\hbar^2}{m}\epsilon+\mathcal{O}(\epsilon^2).
\end{align}
We will also need the total energy in the high-velocity regime, that may be expressed as \cite{jones_motions_1982}
\begin{align}
	\label{Etotlinear}
	E_\text{tot}=\frac{8\pi}{3}\frac{n_0\hbar^2}{m}\epsilon+\mathcal{O}(\epsilon^2).
\end{align}
In the high-velocity limit the total energy is twice the interaction energy.

\begin{figure}
	\centering
	\includegraphics[width=1\linewidth]{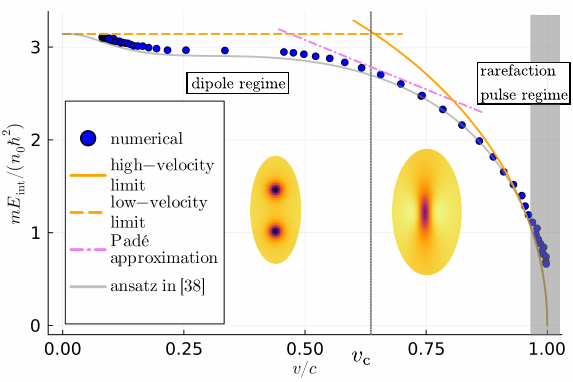}
	\caption{Interaction energy as a function of the velocity. The blue dots stem from numerical calculations (see appendix \ref{Simulation}), the orange dashed line is the interaction energy in the low-velocity limit \cite{kuznetsov_instability_1995} and the orange solid line is the high-velocity result (\ref{Eintlinear}). The grey dots signify the critical velocity $v_\text{c}\simeq 0.64c$ at which the annihilation occurs \cite{jones_motions_1986}. The violet dash-dotted line is derived using Padé approximations for the density as discussed in section \ref{Pade}. We observe good agreement to the dispersion relation ansatz approximation found in \cite{smirnov_dynamics_2012} (grey solid line). The early decline near $v\sim 0.1c$ is due to the weak mutual distortion of independent vortex cores when  $d_\textrm{v}\lesssim 14\xi$. The velocity was extracted by calculating the momentum and interaction energy and employing (\ref{Integral-relation2}). For high velocities $v\sim c$ the interaction energy and momentum become sensitive to small changes in velocity. Additionaly, JRS become increasingly flat making the estimation of these quantities sensitive to collisions with sound waves. These effects lead to an overestimation of the velocity in the region marked with a grey band. For smaller interaction energies $mE_\text{int}/(n_0\hbar)\lesssim 0.5$, the soliton reaches an extension comparable to the size of the grid used in the simulation.}
	\label{fig:eintvsv}
\end{figure}
Figure \ref{fig:eintvsv} shows the interaction energy as a function of the velocity. Indeed the interaction energy is essentially constant until the annihilation, so that the velocity and hence the distance between the vortices captures all the physics in the dipole regime. On the other hand, in the high-velocity limit the interaction energy changes dramatically with minimal changes in velocity. In the regime close to the annihilation, both quantities vary. However, while there is no obvious way to extract the velocity without calculating derivatives of noisy phase data, the interaction energy can be obtained easily from the atomic density.

\subsection{Stochastic Projected Gross-Pitaevskii Theory}
In this section we give an overview of the reservoir theory we use to analyze thermally decay of JRS. We will make use of the SPGPE theory \cite{gardiner_stochastic_2003,bradley_bose-einstein_2008,rooney_stochastic_2012,bradley_low-dimensional_2015}. The formalism gives rise to two damping mechanisms in the GPE. We analyze their respective influence on the stability of rarefaction pulses and vortex dipoles. The mechanisms show a qualitatively different behaviour, implying the possibility to easily differentiate between them in experiment.

In the SPGPE reservoir theory, the atomic cloud is divided into a coherent low-energy region and an incoherent high-energy region. The low-energy region is treated dynamically and assumed to consist of highly occupied states (mean occupation number of at least the order one), and the high-energy region is assumed to be thermally distributed. Including both damping terms in the GPE leads to the equation of motion \cite{blakie_dynamics_2008}
\begin{subequations}
	\label{SPGPE}
	\begin{align}
		i\hbar\partial_t\psi(\textbf{r})=&(\mathcal{L}_\text{GP}-\mu)\psi(\textbf{r})+i\hbar\partial_t\psi(\textbf{r})|_\gamma+i\hbar\partial_t\psi(\textbf{r})|_\varepsilon,	\label{hamil}
  \end{align}
  where
  \begin{align}
		i\hbar\partial_t\psi(\textbf{r})|_\gamma&=-i\gamma(\mathcal{L}_\text{GP}-\mu)\psi(\textbf{r})\text{ and}\label{ndampdef}\\
		i\hbar\partial_t\psi(\textbf{r})|_\varepsilon&=-\hbar\int d^2\textbf{r}'\varepsilon(\textbf{r}-\textbf{r}')\nabla'\cdot \textbf{j}(\textbf{r}')\psi(\textbf{r}).
		\label{edampdef}
\end{align}
\end{subequations}
$\textbf{j}$ is the current density
\begin{align}
	\textbf{j}=\frac{\hbar}{m}\text{Im}\left\{\psi^*\nabla\psi\right\}.
\end{align}

The first mechanism, (\ref{ndampdef}), so called number damping, is characterised by the dimensionless damping strength~\cite{bradley_bose-einstein_2008}
\begin{align}\label{gamdef}
	\gamma&=\frac{8a_\text{s}^2}{\lambda_\text{th}^2}e^{\beta\mu_{3D}}\sum_{j=1}^\infty z^j\Phi\left[z,1,j\right]^2,
\end{align}
where $z=e^{\beta(\mu_{3D}-2\epsilon_\text{cut})}$, $\beta=1/(k_BT)$, and $\Phi[z,s,\alpha]=\sum_{n=0}^\infty z^n/(n+\alpha)^s$ the Lerch transcendent and the cutoff $\epsilon_\text{cut}$ is the energy above which all modes are treated as a thermal reservoir. This cutoff has typically to be choosen as $\epsilon_\text{cut}\sim2\mu_{3D}-3\mu_{3D}$, where a small change in its choice has a small effect on $\gamma$ \cite{gardiner_stochastic_2003}\footnote{If $\epsilon_\text{cut}$ is varied by $\alpha \epsilon_\text{cut}$ with $\alpha\ll1$, $\gamma$ changes in leading order by less than $-\alpha\beta\epsilon_\text{cut}[(N_\text{cut}+1)^2+1/(1-z)]\gamma$, where $N_\text{cut}$, $\beta\epsilon_\text{cut}$ are of order one and $z$ is typically not close to one.}. As our results are linear in $\gamma$, the cutoff dependence of all (number) damping terms derived in the following are fully characterised by $\gamma$.The number damping process involves particle transfer from the reservoir to the coherent region below the cutoff, driving condensation. It describes scattering events in which one of the reservoir atoms loses enough energy to reach the coherent region. Since $\gamma\ll 1$ it is formally equivalent to the damped Gross-Pitaevskii equation (DGPE) \cite{pitaevskii_phenomenological_1959,choi_phenomenological_1998,madarassy_vortex_2008,sergeev_mutual_2023}, that is often used to describe dissipative BEC dynamics. In the following, we will refer to the GPE including the number damping process (\ref{ndampdef}) as $\gamma$GPE and mark a quantity $A$ evolving according to this equation as $A|_\gamma$.

The second decay process, (\ref{edampdef}), so-called energy damping, is characterised by the integral kernel $\varepsilon(\mathbf{r})$. It results from the scattering between an atom in the coherent region with one in the incoherent region. The former transfers energy to the latter, thereby damping excitations in the BEC. In contrast to number damping, this process is number conserving. 
The integral kernel in (\ref{edampdef}) characterising energy damping in two dimensions is defined by its Fourier-transform~\cite{bradley_low-dimensional_2015}
\begin{align}
	\label{kernel}
	\tilde{\varepsilon}(\mathbf{k})&=\int d^2\mathbf{r}e^{-i\mathbf{k}\cdot\mathbf{r}}\;\varepsilon(\mathbf{r})=8a_\text{s}^2N_\text{cut}F\left(\left|\frac{l_z\mathbf{k}}{2}\right|^2\right).
\end{align}
Here $F(z)\equiv e^zK_0(z)$ denotes the scaled modified Bessel-function of the second kind and $N_\text{cut}=(\exp[\beta (\epsilon_\text{cut}-\mu_{3D})]-1)^{-1}$ is the thermal population of modes at the cutoff energy. This later quantity contains the full dependence on the cutoff energy in the energy damping term. As our analysis is linear in the damping, the entire cutoff dependence of all (energy) damping terms derived in the following sections is given by $N_\text{cut}$, which in the validity regime of the SPGPE varies only slightly if the cutoff energy is varied slightly\cite{gardiner_stochastic_2003}\footnote{f $\epsilon_\text{cut}$ is varied by $\alpha \epsilon_\text{cut}$ with $\alpha\ll1$, $N_\text{cut}$ changes in leading order by $-\alpha\beta\epsilon_\text{cut}(1+N_\text{cut})N_\text{cut}$, where $N_\text{cut},\ \beta\epsilon_\text{cut}$ are of order one.\cite{gardiner_stochastic_2003}}. We are working with a reduction of the SPGPE into two dimensions~\cite{bradley_low-dimensional_2015}, valid as long as the trapping in the third direction is strong enough. More precisely, the oscillator length $l_z$ should be on the order of a (few) healing length $\xi$ \cite{rooney_suppression_2011}, which is also where a dimensional reduction of the GPE is valid. In the following, we will refer to the GPE including the energy damping process (\ref{edampdef}) as $\varepsilon$GPE and mark a quantitiy $A$ evolving according to this equation with $A|_\varepsilon$.

The complete SPGPE also contains an explicit projection on the (low-energy) coherent states and a noise term for each process, both neglected in (\ref{SPGPE}). In homogenous systems the projector prohibits occupation of high momenta states in the wavefunction $\psi$. However, JRS are quite smooth and are expected to be adequately captured by the low momenta states with no significant dependence on the high momentum states. Therefore, we will neglect the projection throughout this work. Moreover, as long as the excitation energy of the JRS is large compared to the thermal energy scale $k_\text{B}T$, we can expect the damping to dominate over the noise so that the latter can be ignored in the description (see appendix \ref{Noise} for a brief discussion of the importance of noise). Throughout, we will assume the temperature to be sufficiently small for this condition to be satisfied, and we therefore neglect the noise as a higher order correction to the dynamics.

\section{Analytical Results}
\label{weakdamping}

This section is devoted to the analytical study of decaying JRS. We derive general damping rates in terms of the momentum $P$ (equations (\ref{PN}) and (\ref{PE})). For vortex dipoles the momentum is up to prefactors given by the distance $d_\text{v}$ between the vortices, the characterising property in this regime. For fast moving JRS, the momentum becomes proportional to the interaction energy $E_\text{int}$, which we identify as the characterising property for that case.
In order to study the damping processes, we employ a perturbed Euler-Lagrange equation (pELE) approach~\cite{kivshar_lagrangian_1995}\footnote{In \cite{kivshar_lagrangian_1995} the pELE as well as the Lagrangian are derived in one dimension. However, Jones-Roberts solitons that must obey (\ref{soliton}), the extension to higher dimensions follows straightforwardly.}. For GPE evolution subject to a weak perturbation in the form 
\begin{align}
	i\hbar\partial_t\psi&=({\cal L}_{\textrm{GP}}-\mu)\psi + {\cal U}\psi,
\end{align}
the soliton dynamics is governed by the pELE 
\begin{align}
	\label{pELE}
	\frac{\partial L}{\partial x_\text{s}}-\frac{d}{dt}\frac{\partial L}{\partial \dot{x}_\text{s}}=2\Re\left\{ \int d^2\textbf{r}(\mathcal{U}\psi)^*\frac{\partial \psi(\textbf{r})}{\partial x_\text{s}}\right\},
\end{align}
where $x_\text{s}$ is again the position of the JRS on the $x$-axis. For any situations where the integral is analytically tractable, the effect of a particular damping mechanism can be evaluated. 

As in the case of vanishing damping (\ref{ELE}), since $L$ does not depend on the position of the JRS, we recover for the left hand side of (\ref{pELE}) the negative time derivative of the momentum, $-\dot{P}$.

We consider two independent damping processes stemming from interactions described by the SPGPE open quantum systems theory of BEC \cite{gardiner_stochastic_2003,bradley_bose-einstein_2008,rooney_stochastic_2012,bradley_low-dimensional_2015}: number damping, that accounts for scattering of atoms in the thermal cloud leading to condensate growth, and energy damping, describing scattering events between atoms in the thermal cloud and in the coherent band. We start with the former.

\subsection{Number Damping}
\label{Number}
Before we can evaluate the right hand side of the perturbed Euler-Lagrange equation (\ref{pELE}), we have to specify the perturbation $\mathcal{U}\psi$. In the case of evolution according to the $\gamma$GPE the perturbation can be identified from (\ref{ndampdef})
\begin{align}\label{Upsi}
	\mathcal{U}\psi=-i\gamma(\mathcal{L}_\text{GP}-\mu)\psi.
\end{align}

We assume $\psi$ at any time to be of the form of a JRS (\ref{solitonic}) with (time dependent) velocity $v$. Additionally, due to Eqs. (\ref{Upsi}) and (\ref{stationary}) we have
\begin{align}
	\mathcal{U}\psi=-\gamma v\hbar\partial_x\psi.
\end{align}
Noting $\partial\psi/\partial x_\text{s}=-\partial\psi/\partial x$ equation (\ref{pELE}) becomes
\begin{align}
	\label{numberP}
	-\frac{dP}{dt}\bigg|_\gamma=2\gamma v\hbar\int d^2\textbf{r}|\partial_x\psi|^2.
\end{align}
Using (\ref{Integral-relation1}) we obtain
\begin{align}
	\label{PN}
	\frac{dP}{dt}\bigg|_\gamma=-2\frac{\gamma mv}{\hbar}E_\text{tot}.
\end{align}
The number damping induced decay rate for the momentum is thus proportional to the product of the total energy and the velocity. These results hold under the assumption that $\psi$ stays in the form of a JRS at all times\footnote{See appendix \ref{Validity} for a discussion of the breakdown of this condition. Note that the argumentation is also valid for one and three dimensional solitons.}. Employing the relation (\ref{Integral-relation2}) we can rewrite this result to
\begin{align}
	\label{ndamp}
	\frac{d}{dt}\frac{E_\text{int}}{v}\bigg|_\gamma=-\frac{\gamma mv}{\hbar}E_\text{tot},
\end{align}
a form suitable for describing both slow and fast JRS.
First, we consider the low-velocity limit. For two far separated and thus only weakly interacting vortices the interaction energy $E_\text{int}$ is constant. Hence, we can neglect its time derivative  in (\ref{ndamp}). Employing (\ref{vd}), (\ref{Etotd}) and (\ref{Eintd}) we can state the decay in terms of the distance between the vortices $d_\text{v}$
\begin{align}
	\label{numberd}
	\frac{d}{dt}d_\text{v}\bigg|_\gamma=-2\gamma\frac{\hbar}{m}\frac{1}{d_\text{v}}\ln\left(\alpha\frac{d_\text{v}}{\xi}\right)+\mathcal{O}\left(\frac{1}{d_\text{v}^2}\right).
\end{align}
Number damping hence leads to a logarithmically scaled approach of the vortices as opposed to the pure antiproportional behaviour seen in the literature~\cite{tornkvist_vortex_1997,mehdi_mutual_2023,sergeev_mutual_2023}. We verify the validity of this log correction in section \ref{highlow}.

We now consider the high-velocity limit $v\lesssim c$. In this regime, the time derivative of the velocity $\partial_t v$ becomes negligibly small. To be more precise we consider the behaviour of the reciprocal Lorentz factor $\epsilon=\sqrt{1-(v/c)^2}$. While $\partial_t v$ is of order $\epsilon\partial_t\epsilon$, the interaction energy $E_\text{int}$ is at leading order linear in $\epsilon$ (see (\ref{Eintlinear})). By neglecting the time derivative of the velocity, we hence make a mistake two orders higher in $\epsilon$ than the lowest order in the result will be. Using the observation below (\ref{Etotlinear}) that $E_\text{tot}\simeq 2E_\text{int}$ in this regime, we find
\begin{align}
	\label{numberdamping}
	\begin{split}
		\frac{d E_\text{int}}{dt}\bigg|_\gamma&=-\frac{2\gamma mv^2}{\hbar}E_\text{int}+\mathcal{O}(\epsilon^3)\\
		&=-\frac{\gamma \hbar}{m\xi^2}E_\text{int}+\mathcal{O}(\epsilon^3),
	\end{split}
\end{align}
where we also replaced $v^2$ with $c^2=\hbar^2/(2m^2\xi^2)$ in the last line\footnote{See appendix \ref{higherordercorrections} for the next higher order in interaction energy $E_\text{int}^3$.}. In the high-velocity limit, we thus expect number damping to induce an exponential decay in the interaction energy given by the number damping strength $\gamma$ 
\begin{align}
	E_\text{int}(t)|_\gamma=E_\text{int}(0)\exp\left(-\frac{\gamma \hbar}{m\xi^2}t\right).
\end{align}

The interaction energy is in the lowest order of the reciprocal Lorentz factor $\epsilon$ proportional to the latter, see (\ref{Eintlinear}). Hence, we can express our result in terms of $\epsilon$
\begin{align}
	\label{numberdamp}
	\frac{d\epsilon}{dt}\bigg|_\gamma=-\frac{\gamma\hbar}{m\xi^2}\epsilon+\mathcal{O}(\epsilon^2).
\end{align}
As the depth of the soliton is proportional to $\epsilon^2$ and its extension to $\xi/\epsilon^2$ (see (\ref{rhohighv})), number damping leads it to grow exponentially flatter and shallower.

\subsection{Energy Damping}
\label{Energy}
The SPGPE does not only describe condensate growth in the form of number damping. It also describes the scattering between the atoms in the coherent band with the atoms in the thermal cloud. We can identify the perturbation caused by this so called energy damping from (\ref{edampdef})
\begin{align}
	\label{energydamping}
	\begin{split}
		\mathcal{U}\psi(\textbf{r})&\equiv-\hbar\int d^2\textbf{r}'\varepsilon(\textbf{r}-\textbf{r}')\nabla'\cdot \textbf{j}(\textbf{r}')\psi(\textbf{r})\\
		&=\hbar\int d^2\textbf{r}'\varepsilon(\textbf{r}-\textbf{r}')\partial_tn(\textbf{r}')\psi(\textbf{r}),
	\end{split}
\end{align}
with the kernel $\varepsilon$ as given in (\ref{kernel}). The second equality holds because energy damping does not allow for the exchange of particles between the bands and thus is number-conserving. We are neglecting number damping in this section. If included, the system is no longer number-conserving. However, the change in particle number is of order $\gamma$ and hence usually neglectable in (\ref{energydamping}).

The right hand side of the perturbed Euler-Lagrange equation (\ref{pELE}) becomes thus
\begin{align}
	\begin{split}
		&2\Re\left\{\int d^2\textbf{r}\hbar\int d^2\textbf{r}'\varepsilon(\textbf{r}-\textbf{r}')\partial_tn(\textbf{r}')\psi^*(\textbf{r})\partial_{x_\text{s}}\psi(\textbf{r})\right\}\\
		&=-\hbar\Re\left\{\int d^2\textbf{r}\int d^2\textbf{r}'\varepsilon(\textbf{r}-\textbf{r}')\partial_tn(\textbf{r}')\partial_xn(\textbf{r})\right\}\\
		&=\hbar v\int d^2\textbf{r}\int d^2\textbf{r}'\varepsilon(\textbf{r}-\textbf{r}')\partial_{x'}n(\textbf{r}')\partial_xn(\textbf{r})\\
		&=\hbar v\int \frac{d^2\textbf{k}}{(2\pi)^2}\tilde{\varepsilon}(\textbf{k})\widetilde{\partial_xn}(\textbf{k})\widetilde{\partial_xn}(-\textbf{k}),
	\end{split}
\end{align}
where we used $\partial_t n=-v\partial_xn$ from the second to the third line. The pELE (\ref{pELE}) hence predicts damping according to
\begin{align}
	\label{PE}
	\begin{split}
		\frac{dP}{dt}\bigg|_\varepsilon=&-\hbar v\int \frac{d^2\textbf{k}}{(2\pi)^2}\tilde{\varepsilon}(\textbf{k})\left|\widetilde{\partial_xn}(\textbf{k})\right|^2.\\
	\end{split}
\end{align}
Again, this holds under the assumption that $\psi$ stays a JRS at all times. This yields with (\ref{Integral-relation2})
\begin{align}
	\label{edampi}
	\begin{split}
		\frac{d}{dt}\frac{E_\text{int}}{v}\bigg|_\varepsilon=&-\frac{\hbar v}{2}\int \frac{d^2\textbf{k}}{(2\pi)^2}\tilde{\varepsilon}(\textbf{k})\left|\widetilde{\partial_xn}(\textbf{k})\right|^2.\\
	\end{split}
\end{align}
Interestingly, energy damping seems not to depend directly on the phase and mainly works against strong fluctuations in the density\footnote{Since the Fourier transformed kernel $\tilde{\varepsilon}$ depends only on the absolute value of its argument, we will use scalars as arguments of $\tilde{\varepsilon}$ in the following.}.

We first have a look at the low-velocity limit, which was already analyzed in \cite{mehdi_mutual_2023}. If $|\widetilde{\partial_xn}|^2$ is strongly peaked compared to $\tilde{\varepsilon}$, we can approximate the latter by its value at the maximum of the prior $k_\text{max}$. In \cite{mehdi_mutual_2023} it is argued that this is allowed in the case of widely separated vortices, corresponding for two oppositely charged vortices to the low-velocity limit of a planar Jones-Roberts soliton. Under this assumption, the perturbed Euler-Lagrange equation (\ref{edampi}) becomes
\begin{align}
	\label{edamp}
	\begin{split}
		\frac{d}{dt}\frac{E_\text{int}}{v}\bigg|_\varepsilon\simeq&-\frac{\hbar v}{2}\tilde{\varepsilon}(k_\text{max})\int \frac{d^2\textbf{k}}{(2\pi)^2}\left|\widetilde{\partial_xn}(\textbf{k})\right|^2\\
		=&-\frac{\hbar v}{2}\tilde{\varepsilon}(k_\text{max})\int d^2\textbf{r}\left(\partial_xn(\textbf{r})\right)^2.
	\end{split}
\end{align}
\begin{table*}
	\begin{tabular}{cccc}
		 &Jones-Roberts soliton  & \,\,Vortex dipole\,\, & \,\,Rarefaction pulse\,\, \\
		\hline\hline
		Velocity regime & $0<v< c$&$v\ll c$ &$v\lesssim c$   \\
		Decay rate &$\dot{P}/P$ & $\dot{d}_\text{v}/d_\text{v}$ &  $ \dot{E}_\text{int}/E_\text{int}$ \\
		Number damping & Eq.~(\ref{PN}) & Eq.~(\ref{ndamp}) & Eq.~(\ref{numberdamping}) \\
		Energy damping & Eq.~(\ref{PE}) & Eq.~(\ref{MehdiResult}) & Eq.~(\ref{energydamp}) \\
  \hline
	\end{tabular}
	\caption{JRS damping characterized by the momentum decay mechanism in the different regimes. In the dipole regime the momentum is proportional to the distance between the vortices $d_\text{v}$, so that the normalized momentum decay becomes the normalized decay of this distance. In the rarefaction pulse regime the momentum is proportional to the interaction energy $E_\text{int}$, so that the normalized momentum decay becomes the normalized decay in interaction energy. }
	\label{table:comparison}
\end{table*}
Employing (\ref{vd}) we derive
\begin{align}
	\frac{d}{dt}d_\text{v}\bigg|_\varepsilon\simeq-\frac{\hbar^3}{2m^2} \frac{1}{d_\text{v}}\tilde{\varepsilon}(k_\text{max})\frac{\int d^2\textbf{r}\left(\partial_xn(\textbf{r})\right)^2}{E_\text{int}}.
\end{align}
As the JRS in the low-velocity limit is essentially described by two far separated vortices, the density close to one of the vortices looks like the density of a single vortex $n_\text{vortex}$. The integrals can thus be calculated as twice the integrals using the single vortex density. If we use, as was done in \cite{mehdi_mutual_2023}, a Gaussian ansatz for this density
\begin{align}
	n_\text{vortex}(\textbf{r})=n_0\left(1-\exp\left[-\frac{\textbf{r}^2}{4\xi^2}\right]\right),
\end{align}
we see that $\widetilde{\partial_xn}_\text{vortex}$ is sharply peaked around $1/(\sqrt{2}\xi)$. With this we recover their result (note carefully the differing definition of the healing length by a factor of $\sqrt{2}$) for the damping of a vortex dipole
\begin{align}
	\label{MehdiResult}
	\begin{split}
		\frac{d}{dt}d_\text{v}\bigg|_\varepsilon\simeq&-\frac{\hbar^3}{2m^2} \frac{1}{d_\text{v}}\tilde{\varepsilon}\left(\frac{1}{\sqrt{2}\xi}\right)\frac{n_0^2\pi}{n_0^2g\xi^22\pi}\\
		=&-4n_0a_\text{s}^2N_\text{cut}F\left(\frac{l_z^2}{8\xi^2}\right)\frac{\hbar}{md_\text{v}}.
	\end{split}
\end{align}

We now analyze the high-velocity limit, long after the collapse of two vortices into a rarefaction pulse. We are searching for an expression of the damping in lowest order of the reciprocal Lorentz factor $\epsilon$ (and thus, due to (\ref{Eintlinear}), in the interaction energy). First, we can approximate the velocity $v$ by the speed of sound $c$ and hence neglect the time derivative of the velocity appearing in the pELE (\ref{edampi}) (cf. section \ref{Number}). Upon calculating the integral appearing in (\ref{edampi}) we then obtain\footnote{We consider the limit $\epsilon\ll1$. However, deriving this rate requires the usually slightly stronger condition $\epsilon\ll\xi/l_z$.} (see appendix \ref{partrho})
\begin{align}
	\label{energydamp}
	\begin{split}
		\frac{dE_\text{int}}{dt}\bigg|_\varepsilon&=\frac{6}{\pi^2}\frac{ma_\text{s}^2}{\hbar^3n_0\xi^2}N_\text{cut}E_\text{int}^3\left[2\ln\left(\frac{3mE_\text{int}}{4\pi n_0\hbar^2}\frac{l_z}{\xi}\right)+C_\text{E}\right]\\
  &+\mathcal{O}(\epsilon^4)
	\end{split}
\end{align}
and expressed in the reciprocal Lorentz factor
\begin{align}
	\label{epsilonenergydamp}
	\begin{split}
		\frac{d\epsilon}{dt}\bigg|_\varepsilon=&\frac{32}{3}n_0a_\text{s}^2\frac{\hbar}{m\xi^2}N_\text{cut}\epsilon^3\left[2\ln\left(\epsilon\frac{l_z}{\xi}\right)+C_\text{E}\right]+\mathcal{O}(\epsilon^4).
	\end{split}
\end{align}
Both results are valid in lowest order of the reciprocal Lorentz factor. We introduced the constant
\begin{align}
	C_\text{E}=-\ln(12)-\gamma_\text{Euler}+\frac{5}{2}=-0.562...,
\end{align}	
with $\gamma_\text{Euler}=0.577...$ the Euler-Mascheroni constant, in order to write our results more compactly. Remarkably, in lowest order of $\epsilon$ the thickness of the atomic cloud $l_z$ is unimportant. The rate with which the JRS is damped depends hence on the fraction of the area of the condensate in which atoms are susceptible to scattering with thermal atoms $n_0\sigma_\text{s}$ (with $\sigma_\text{s}=4\pi a_\text{s}^2$ the scattering cross section) and the number of particles in the low-energy state of the thermal cloud $N_\text{cut}$.

Interestingly, energy damping shows a qualitatively distinct behaviour than the exponential number damping (\ref{numberdamping}). Especially, the normalized damping rate features an extremum and falls towards zero for small interaction energies, while the normalized number damping rate does not feature any local extrema and maximizes for small interaction energies. The dominant decay mechanism is hence identifiable by qualitatively studying the diminishing interaction energy.

\subsection{Comparison between the Damping Mechanisms}
\label{comparison}
In the previous sections, we obtained results for the two damping mechanisms occurring in BECs according to the $\gamma$GPE and $\varepsilon$GPE. Table \ref{table:comparison} summarizes the decay rates of the momentum in the two regimes and under the two damping mechanisms studied. We now demonstrate that the relevant process in the decay of a planar Jones-Roberts soliton in an experiment will usually be energy damping. 

In the case of a dipole, the dominance of energy damping for experimental parameters was already discussed in \cite{mehdi_mutual_2023}. However, we extend this discussion by giving bounds on the respective dominance of number and energy damping.

From (\ref{numberd}) and (\ref{MehdiResult}) we obtain
\begin{align}
	\label{comlow}
	\frac{\dot{d}|_\varepsilon}{\dot{d}|_\gamma}=\frac{2n_0a_\text{s}^2N_\text{cut}F(l_z^2/[8\xi^2])}{\gamma\ln(\alpha d_\text{v}/\xi)}.
\end{align}
Energy damping is the relevant process in the PV regime if the right hand side is large against one, otherwise number damping is dominant. Using the results from appendix \ref{numberstrength} together with representative parameters for current experiments we can write these conditions for the relevant damping mechanism in terms of conditions on the two-dimensional phase-space density.

First we use the upper bound in (\ref{gammabounds})
\begin{align}
	\label{gammakleiner}
	\gamma<8N_\text{cut}^2\frac{a_\text{s}^2}{\lambda_\text{th}^2},
\end{align}
where $\lambda_\text{th}=\sqrt{2\pi\hbar^2/(mk_\text{B}T)}$ is the thermal de Broglie wave-length. For the gas to be handled as two dimensional its extension in the $z$-direction $l_z$ should not be much larger than a healing length $\xi$ \cite{bradley_low-dimensional_2015}. Moreover, to treat $\psi$ as a classical field requires $N_\text{cut}$ to be of order one \cite{gardiner_stochastic_2003}. Due to limitations in their size, current experiments can hardly support dipoles with a distance much larger than $100\xi$ \cite{moon_thermal_2015,meyer_observation_2017,kwon_sound_2021}. These considerations imply that given
\begin{align}
	\label{edom1}
	n_0\lambda_\text{th}^2>10
\end{align}
the right hand side of equation (\ref{comlow}) is larger than one. Hence, $\dot{d}|_\varepsilon/\dot{d}|_\gamma> 1$ is guaranteed and an experiment featuring a vortex dipole is dominated by energy damping.

On the other hand, to understand when number damping is important we can use the lower bound from (\ref{gammabounds})
\begin{align}
	\label{gammagrosser}
	\gamma>8\frac{N_\text{cut}^2}{(N_\text{cut}+1)^2}\frac{a_\text{s}^2}{\lambda_\text{th}^2},
\end{align}
again taking $N_\text{cut}$ of order one and $l_z$ on the order of a healing length. Further, we can not expect the PV limit and hence our calculations to hold anymore, when the vortices are only a few healing lengths apart. We thus obtain the condition
\begin{align}
	\label{ndom1}
	n_0\lambda_\text{th}^2<1,
\end{align}
guaranteeing that number damping will be the relevant process in the approach of the vortices. It should be noted that in a degenerate Bose gas this condition will usually not be satisfied and the inclusion of energy damping can be expected to be crucial. Number damping is likely the dominant process for close to the boundary of a trapped condensate, where the density is low.

\begin{figure}
	\centering
	\includegraphics[width=1\linewidth]{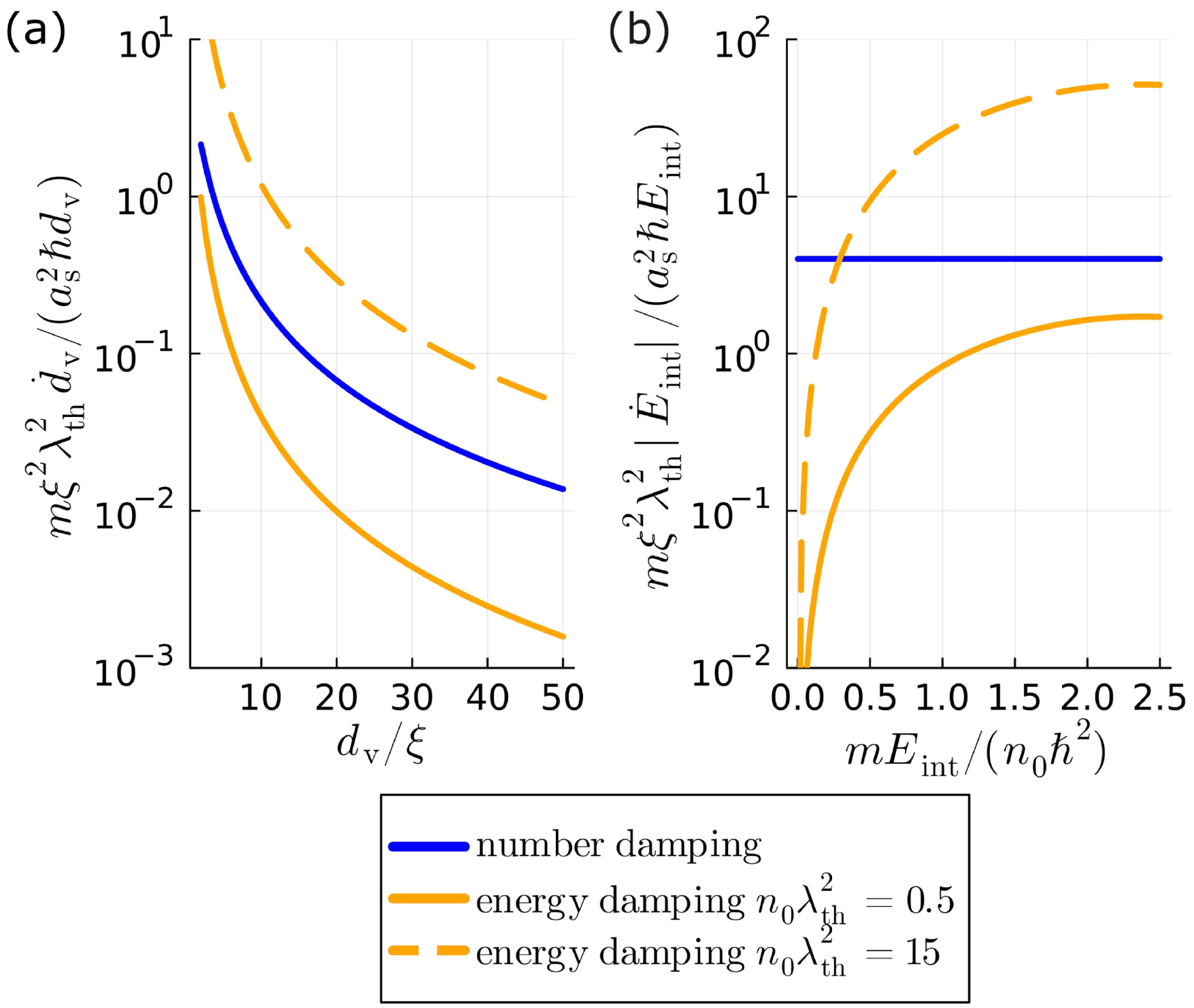}
	\caption{Normalized decay of the momentum under number damping (blue) compared with its decay under energy damping in the low (orange, solid) and high (orange, dashed) 2D phase space density regime. (a) shows the analytical results (\ref{numberd}), (\ref{MehdiResult}) for the damping of the vortex distance of a dipole, (b) the results (\ref{numberdamping}), (\ref{energydamp}) concerning the decay of the interaction energy of a rarefaction pulse. As in both cases the normalized decay of the momentum is plotted against the momentum (this is because the momentum is proportional to $d_\text{v}$ in the low-velocity limit and to $E_\text{int}$ in the high-velocity limit), a direct comparison of the damping strengths is possible. A cutoff is chosen so that $N_\text{cut}=1$. The number damping strength is set as $\gamma=4a_\text{s}^2/\lambda_\text{th}^2$, corresponding to $\beta\mu_{3D}\simeq0.58$ (other choices of the temperature and chemical potential can change $\gamma$ only by less than a factor of 2). The thickness of the atomic cloud is set to $l_z=\sqrt{2}\xi$. In the low density regime number damping is the dominant process for all interaction energies for which our analytical results can be expected to lie in the right order of magnitude. On the other hand, in the high density limit, energy damping is dominant for all but extremely large vortex distances or small interaction energies. Hence, it is the relevant process for JRS observable in experiment. As the normalized number damping strength in the rarefaction pulse regime is constant while the normalized energy damping strength varies strongly with interaction energy, the relevant process should be easily identifiable.}
	\label{fig:highvslowdensity}
\end{figure}
We now focus on the rarefaction pulse. For this, we calculate the quotient of the damping rates for energy (\ref{energydamp}) and number (\ref{numberdamping}) damping:
\begin{align}
	\label{comhigh}
	\begin{split}
		\frac{\dot{E}_\text{int}|_\varepsilon}{\dot{E}_\text{int}|_\gamma}=&-\frac{6}{\pi^2}\frac{m^2a_\text{s}^2}{\hbar^4n_0}\frac{N_\text{cut}}{\gamma}E_\text{int}^2\left[2\ln\left(\frac{3mE_\text{int}}{4\pi n_0\hbar^2}\frac{l_z}{\xi}\right)+C_\text{E}\right].
	\end{split}
\end{align}

Using again (\ref{gammakleiner}) we can write this in terms of the reciprocal Lorentz factor
\begin{align}
	\begin{split}
		\frac{\dot{\epsilon}|_\varepsilon}{\dot{\epsilon}|_\gamma}>&-\frac{4}{3N_\text{cut}}\epsilon^2\left[2\ln\left(\epsilon\frac{l_z}{\xi}\right)+C_\text{E}\right]n_0\lambda_\text{th}^2.
	\end{split}
\end{align}
For experimentally observable cases the reciprocal Lorentz factor $\epsilon$ should not be much smaller than $0.2$ (then the depth of the JRS is $\sim n_0/10$ and its spatial extension of the order of hundred healing lengths). As argued before, $N_\text{cut}$ is of order one and $l_z$ on the order of a healing length. Therefore, energy damping is the relevant process provided
\begin{align}
	\label{edom2}
	n_0\lambda_\text{th}^2>10.
\end{align}
Thus, for high densities and low temperatures, number damping can be safely neglected in the decay of a rarefaction pulse.

Surprisingly, the bound for the dominance of energy damping is the same as in the vortex dipole case (\ref{edom1}). This is a coincidence stemming from the spatial extent of current experiments.

On the other hand, (\ref{gammagrosser}) leads to
\begin{align}
	\begin{split}
		\frac{\dot{\epsilon}|_\varepsilon}{\dot{\epsilon}|_\gamma}<&-\frac{4(N_\text{cut}+1)^2}{3N_\text{cut}}\epsilon^2\left[2\ln\left(\epsilon\frac{l_z}{\xi}\right)+C_\text{E}\right]n_0\lambda_\text{th}^2.
	\end{split}
\end{align}
This suggests that number damping is the relevant process and energy damping can be neglected if
\begin{align}
	\label{ndom2}
	n_0\lambda_\text{th}^2<1.
\end{align}

The bound for the dominance of number damping in the decay of a rarefaction pulse is the same as in the vortex dipole case (\ref{ndom1}). Here, however, the dominance of number damping for a 2D phase space density smaller than one seems to be a more general result, as the bound is determined by the behaviour towards the intermediate velocity regime in both cases. As already noted in the dipole case, this condition usually will not be met, requiring the inclusion of energy damping.\footnote{While in 2D we concluded that the phase space density $n_0\lambda_\text{th}^2$ determines the relative strength between energy and number damping, we can expect that in 1D and 3D the healing length comes into play to: As (\ref{PN}) and (\ref{PE}) are also valid in 1D and 3D (for solitons and 3D JRS, respectively),  $n_{1\text{D}}\lambda_\text{th}^2/\xi_{1\text{D}}$ and $n_{3\text{D}}\lambda_\text{th}^2\xi_{3\text{D}}$ (with $n_{j\text{D}}$, $\xi_{j\text{D}}$ the $j$D density and healing length) determine the relative strength in 1D and 3D, respectively.}

Figure \ref{fig:highvslowdensity} compares our analytical results for energy (\ref{MehdiResult}), (\ref{energydamp}) and number damping (\ref{numberd}), (\ref{numberdamping}), illustrating the respective dominances in the two different regimes. It also emphasises the strong qualitative contrast in the rarefaction pulse regime between the exponential number damping and energy damping.

As number damping works exponentially in the high-velocity and logarithmically stronger than energy damping in the low-velocity limit, for small enough interaction energy $E_\text{int}$ or large enough distance between the vortices $d_\text{v}$, respectively, it will eventually become dominant in either case. It appears that energy damping is relevant for small scale excitations, while number damping dominates in the decay of large scale structures. However, it can be expected that in the case of a JRS the size required for number damping to become dominant is usually beyond the scale of current experiments.

\subsection{Estimate of the Lifetime of a JRS}
\label{lifetime}
We want to give an estimate of the lifetime $\Delta t$ of a JRS. An estimate for the lifetime of a dipole was already given in \cite{mehdi_mutual_2023}. Hence, we will focus on how long the rarefaction pulse formed at the collapse of the dipole may persist. However, we do not have an analytical description of the early decay. As will be discussed in section \ref{vortexannihilation}, we can expect our results from the previous sections to hold after the interaction energy reaches roughly $E_0=1.5n_0\hbar^2/m$, so that we will start from there. To get an estimate, we take for the other endpoint the moment in which its interaction energy has fallen to half of this value. For a decay of the interaction energy according to
\begin{align}
	\frac{d E_\text{int}}{dt}=-f(E_\text{int}),
\end{align}
we obtain a lifetime
\begin{align}
	\Delta t=\int_{E_0/2}^{E_0}\frac{dE_\text{int}}{f(E_\text{int})}.
\end{align}
Thus, we derive a lifetime in the case of pure number damping (\ref{numberdamping})
\begin{align}
	\Delta t|_\gamma=\ln(2)\frac{m\xi^2}{\hbar\gamma}
\end{align}
and for energy damping (\ref{energydamp}) (using $\epsilon=3mE/[4\pi n_0\hbar^2]$)
\begin{align}
	\begin{split}
		\Delta t|_\varepsilon=&\frac{\pi^2}{6}\frac{n_0\hbar^3\xi^2}{ma_\text{s}^2N_\text{cut}}\\
		&\times\int_{E_0/2}^{E_0}\frac{dE_\text{int}}{E_\text{int}^{3}}\frac{1}{\left[2\ln\left(\frac{3mE_\text{int}}{4\pi n_0\hbar^2}\frac{l_z}{\xi}\right)+C_E\right]}\\
		=&\frac{3}{32}\frac{m\xi^2}{n_0a_s^2\hbar N_\text{cut}}\int_{0.18}^{0.36}\frac{d\epsilon}{\epsilon^3}\frac{1}{\left[2\ln\left(\epsilon\frac{l_z}{\xi}\right)+C_E\right]}.
	\end{split}
\end{align}

For experiment \cite{meyer_observation_2017} this yields a lifetime of $\Delta t=125$ms (for $N_\text{cut}=1$), consistent with the observed stability over $40$ms. However, this value should be seen more as an order of magnitude estimate as we ignored the inhomogeneity due to the trap as well as treating other properties of the experiment somewhat superficially (see appendix \ref{expparameters}).

\section{Simulations}
\label{numver}
In this section we perform numerical simulations of the damping process to test our predictions for high and low velocities, and explore the intermediate regime. If not otherwise stated we employ a $565\times 565$ healing length domain with $1024\times1024$ points, a time step of $0.03\hbar/\mu$ and periodic boundary conditions. We note that both density and phase fall off to the homogenous background density $n_0$ and zero, respectively. Hence, the exact choice of boundary conditions does not affect the decay as long as the JRS keeps its distance from the boundary.

Exact JRS solutions are numerically challenging to obtain. Here we create JRS at different velocities by starting from a vortex dipole initial condition and imposing weak damping. The solution of the $\varepsilon$GPE adiabatically follows translating ground state, creating a family of stationary state solutions with a range of velocities at fixed atom number (see appendix \ref{Simulation}). Due to the finite numerical domain this procedure also excites weak acoustic artifacts arising from the slow soliton decay and associated release of energy. To extract only the effects of damping and remove the boundary artifacts we subtracted the results of simulations without damping from our damped GPE ($\gamma$GPE and $\varepsilon$GPE, respectively) results (see appendix \ref{simulationcorrection} for details). This proceedure removes the artifacts very effectively, allowing us to concentrate on the role of each damping mechanism. 

For high velocities/low interaction energies the JRS gains an extension that is comparable to the grid size, leading to large numerical errors. However, our choice of grid is sufficient to describe JRS scales in current experiments~\cite{meyer_observation_2017,kwon_sound_2021}.

\subsection{High and Low Velocities}\label{highlow}
In this section we numerically test our analytical findings from section \ref{weakdamping}.

\begin{figure}
	\centering
	\includegraphics[width=1\linewidth]{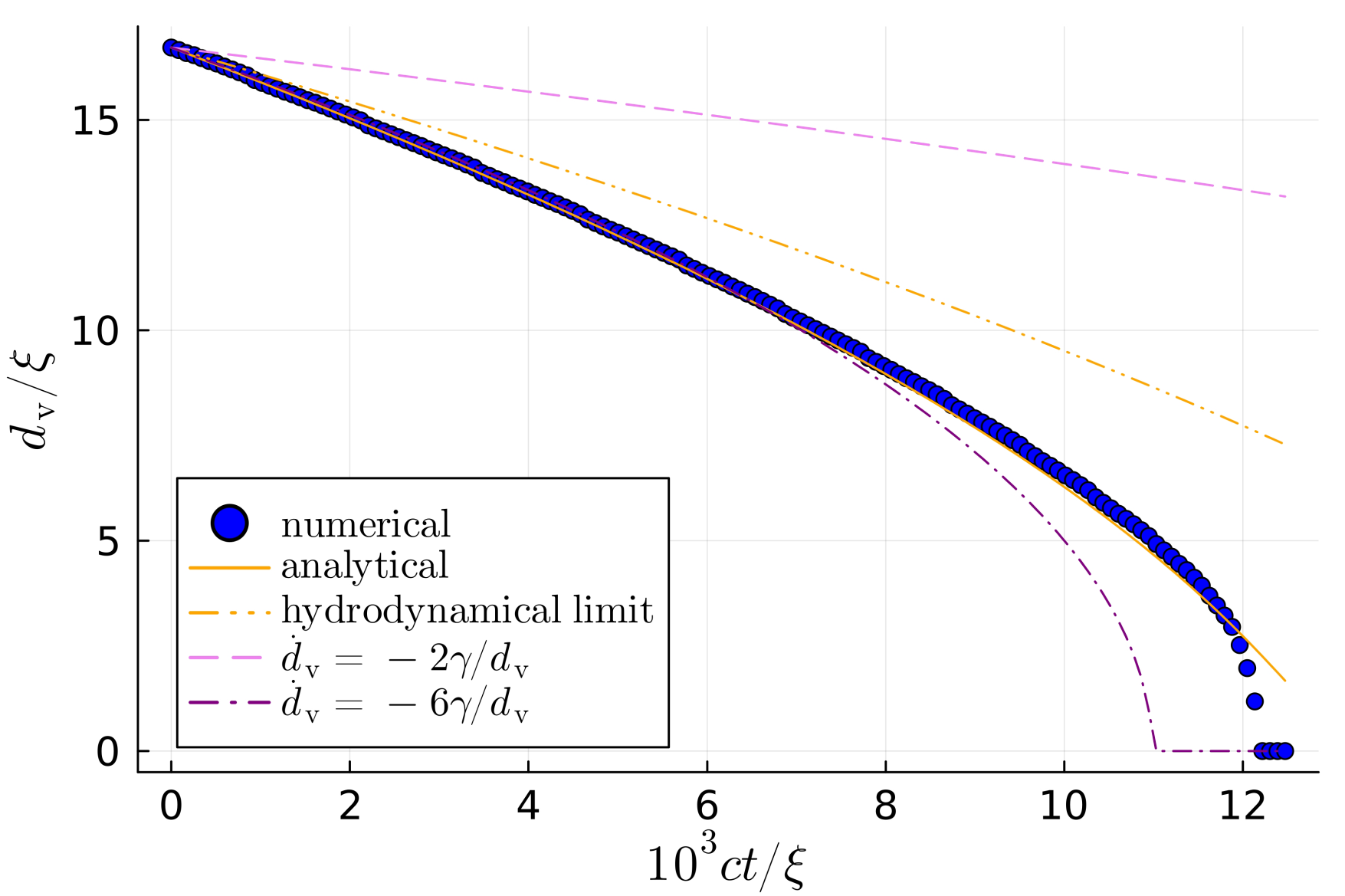}
	\caption{Distance between two vortices against time under evolution according to the $\gamma$GPE with a damping rate $\gamma=0.003$. The blue dots stem from a numerical simulation, where the positions of the two vortices were determined as the global minima of the density in the upper and lower half plane. The orange solid line is the analytical result (\ref{numberd}) and the violet dashed line corresponds to antiproportional behaviour of the damping rate as suggested in \cite{tornkvist_vortex_1997,mehdi_mutual_2023}. The logarithmic correction shows much better agreement with the numerics than the antiproportional damping rate previously claimed. Note that using a phenomenological antiproportional damping rate capturing the early decay correctly (purple dash-dotted line) predicts to strong damping in the long term, thereby demonstrating the necessity of the logarithmic correction. On the other hand, considering only the hydrodynamic limit (corresponding to $\alpha=1/\sqrt{2}$, orange dash-dot-dotted line) predicts considerably to weak damping. Hence, the inclusion of the interaction energy and quantum pressure in the total energy is crucial to gain an accurate description.}
	\label{fig:numberdampingdipole}
\end{figure}
We start with verifying the damping rate found for the decay of a vortex dipole in the $\gamma$GPE (\ref{numberd}). Figure \ref{fig:numberdampingdipole} shows this result together with numerical data and the previously claimed antiproportional ($\dot{d}_\text{v}=-2\gamma/d_\text{v}$) decay. The inclusion of the logarithmic correction shows a much better agreement to the numerics than antiproportional damping. Note that no choice of the prefactor in the antiproportional damping can describe the numerical result over a long time, indicating the importance of the logarithmic scaling. Moreover, neglecting the interaction energy and quantum pressure in the total energy of the dipole (corresponding to setting $\alpha=1/\sqrt{2}$) predicts considerably smaller damping. Thus, their inclusion is also relevant in gaining an adequate description.

\begin{figure}[tb]
	\centering
	\includegraphics[width=1\linewidth]{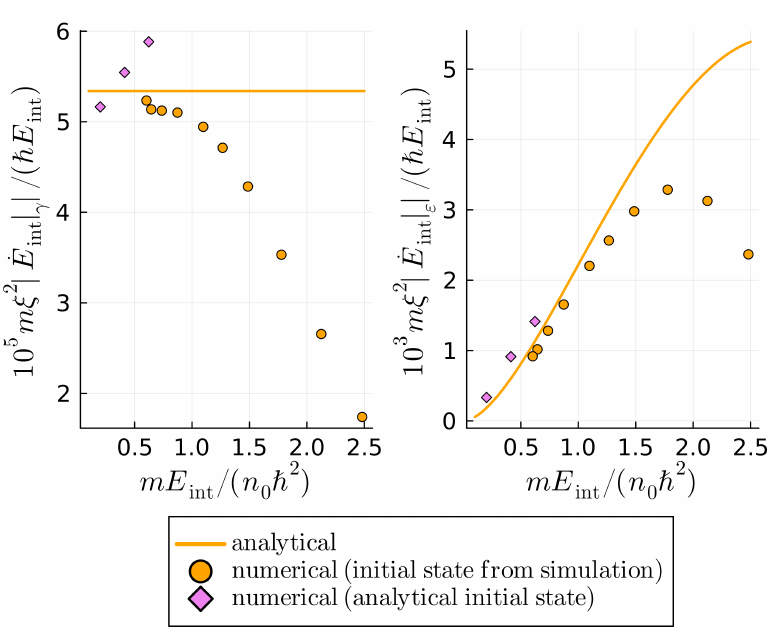}
	\caption{Comparison of the (normalized) analytically derived damping rates (\ref{numberdamping}), (\ref{energydamp}) (orange solid lines) with numerical results (orange dots and violet diamonds). (a) shows the number damping case while (b) presents energy damping induced decay. Our analytical result shows an improving prediction for the damping rate with decreasing interaction energy $E_\text{int}$. The poor agreement for large interaction energies is inevitable, as we work in lowest order of the reciprocal Lorentz factor $\epsilon$ and thus the interaction energy.}
	\label{fig:dEintnumber}
\end{figure}
We now concentrate on the high-velocity regime. In the previous section we identified the interaction energy as a useful quantity in the description of rarefaction pulses. We derived analytical damping rates for the decay of this quantity under number (\ref{numberdamping}) and energy damping (\ref{energydamp}). Figure \ref{fig:dEintnumber} shows these predictions for the damping rates together with numerical results. It demonstrates increasing accuracy of the damping rates derived in the previous section with decreasing interaction energy. Poor agreement for large interaction energies is to be expected, as we work in the lowest order thereof. However, for virtually all interaction energies after the collapse to a rarefaction pulse (occurring at $E_\text{int}\approx2.8n_0\hbar^2/m$, see section \ref{vortexannihilation}), our predictions seem to lie in the correct order of magnitude. The numerics verify the strong qualitative difference between number and energy damping in the high-velocity regime observed in the analytical section. Especially, the normalized energy damping features an extremum, while the normalized number damping does not. For small interaction energies the JRS reaches an extent that is not confined to the grid. On the other hand, the analytical approximation (\ref{rhohighv}, \ref{theta}) becomes valid. We employ it as initial condition on three larger grids. Note, however, that even for small interaction energies the analytical solution is not exact, leading to an accelerated decay in interaction energy. The used parameters stem from the experiment \cite{meyer_observation_2017} (see appendix \ref{expparameters} for details).
\begin{figure}[tb]
	\centering
	\includegraphics[width=1\linewidth]{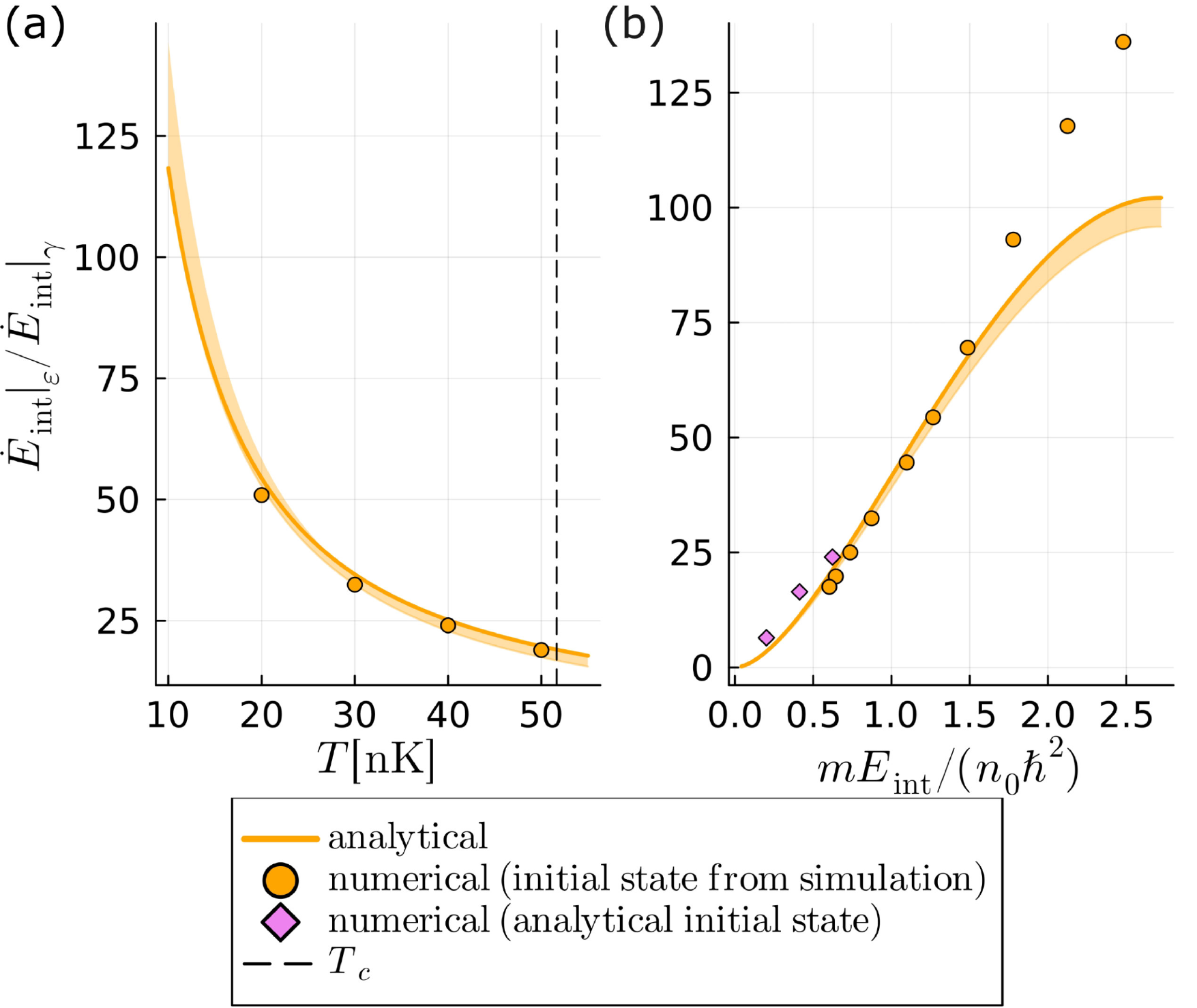}
	\caption{Comparison of the decay of rarefaction pulses due to energy and number damping. The quotient of the energy damping rate $\dot{E}_\text{int}|_\varepsilon$ and the number damping rate $\dot{E}_\text{int}|_\gamma$ is shown for (a) different temperatures $T$ and interaction energy $E_\text{int}=0.87n_0\hbar^2/m$; (b) different interaction energies $E_\text{int}$ at the temperature $T=30$nK. The solid line is the analytical result (\ref{comhigh}), the dots and diamonds stem from numerical calculations. The vertical dashed line marks the critical temperature $T_\text{c}=51.6$nK for an ideal gas in a harmonic trap. Energy damping is the dominant process by one to two orders of magnitude for the used parameters, which correspond to the experimental values in \cite{meyer_observation_2017}, with cutoff $\epsilon_\text{cut}$ chosen such that $N_\text{cut}=1$ (see appendix \ref{expparameters} for details). The shaded region is the analytical estimate for a variation of the cutoff resulting in $1\leq N_\text{cut}\leq 3$.}
	\label{fig:comparison}
\end{figure}
We also derived conditions for the dominance of number (\ref{ndom1}), (\ref{ndom2}) and energy damping (\ref{edom1}), (\ref{edom2}), respectively. We deduced that in an experiment, energy damping can usually be expected to be the dominant mechanism. To verify this numerically we again consider the parameters from the experiment \cite{meyer_observation_2017}, in which rarefaction pulses were observed. The analytically result (\ref{comhigh}) is shown in figure \ref{fig:comparison}, together with numerical results. Energy damping turns out to be the clearly dominant process, being one to two orders of magnitude stronger than number damping for temperatures below the critical temperature and experimentally observable JRS.

\subsection{Intermediate velocities}
\label{vortexannihilation}
In the low-velocity limit, we end up with two widely separated and thus only weakly interacting vortices. This implies that the interaction energy is essentially constant and hence not a useful quantity in this case. Moreover, as already discussed previously, the dynamics in this regime are well described by the PV model \cite{lucas_sound-induced_2014,aref_point_2007,fetter_vortices_2001}. At intermediate velocities, however, two vortices close to each other form a dipole and thereby lose some interaction energy. In the simulation we observe a rapid drop in interaction energy, setting in just before the merger of the two density minima. The interaction energy appears to become the characterizing property just at the annihilation of the dipole and the invalidity of the PV model. This observation motivates our examination of the decay of the interaction energy.

\begin{figure}
	\centering
	\includegraphics[width=1\linewidth]{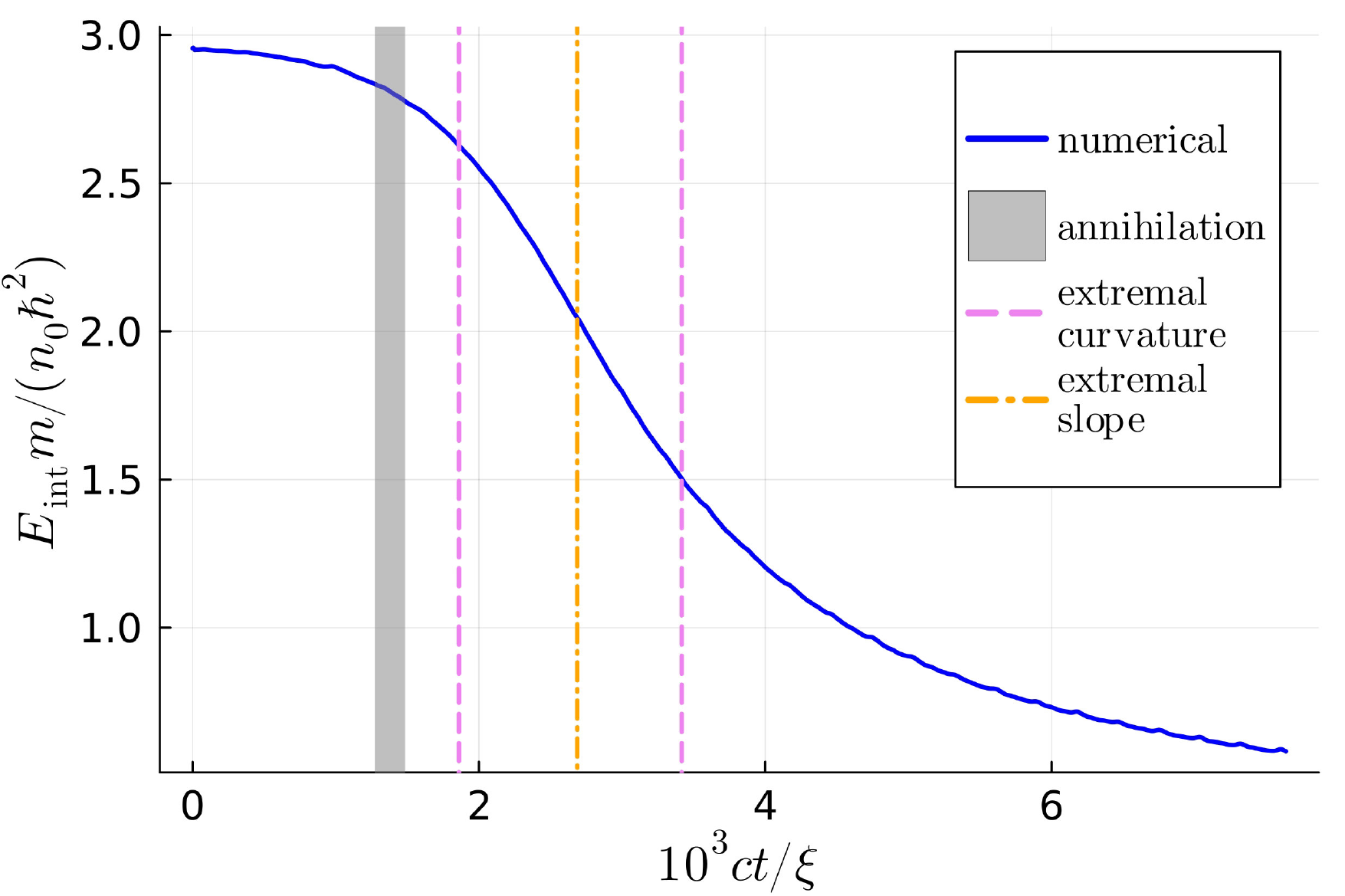}
	\caption{Decay of a vortex dipole: simulation starting with two separated vortices and energy damping with strength $8a_\text{s}^2n_0N_\text{cut}=10^{-3}$ and a thickness of the atomic cloud $l_z=\sqrt{2}\xi$. The grey shaded region marks the disappearance of the dipole. While at its left end, two minima can still be clearly identified in the simulation, at its right end the density of the soliton is everywhere non-zero. The violet dashed lines identify the extrema of the curvature, while the orange dash-dotted line shows the time of extremal slope. These points signify important landmarks in the validity of calculating the energy damping in lowest order in the interaction energy (as was done to derive (\ref{energydamp})): the first extremum of the curvature is reached when the lowest order analytical approximation for the density (\ref{rhohighv}) first becomes non-negative everywhere. At times/interaction energies before the extremum in the slope is reached, the actual damping rate and the analytical result are curved in opposing directions, leading to a rapid worsening of agreement with increasing interaction energy. Beyond the second extremum in the curvature, the analytical result and the actual damping rate are curved in the same direction, both with declining curvature. This finally results in a rapid convergence with decreasing interaction energy.}
	\label{fig:simulationextrema}
\end{figure}
Towards this aim, we run a simulation starting in the PV regime. As we identified energy damping to be the relevant process, we will restrict our discussion to the $\varepsilon$GPE. Figure \ref{fig:simulationextrema} shows the results of the simulation, singling out interesting points in the decay for discussion. While widely separated vortices distort each other only marginally, when they come close to each other they get compressed along the axis perpendicular to the line connecting the two vortices. This process accelerates while they come closer to each other, leading to a rapid decrease in interaction energy. Finally, at an interaction energy of roughly $2.8n_0\hbar^2/m$, the annihilation of the vortices occurs. Two clearly identifiable minima disappear and the density no longer reaches zero. However, after their annihilation, the density is still flattened along the axis that previously connected the two vortices (at the annihilation, the density curvature at the minimum vanishes). The onsetting sharpening along this axis further accelerates the decrease in interaction energy.

Shortly after the disappearance of the dipole, at an interaction energy of roughly $2.6n_0\hbar^2/m$, a clear minimum (this is to say, the density curvature at the minimum gets "large" in the sense of reaching a value comparable with the maximum density curvature for rarefaction pulses at any interaction energy) finally develops. The accompanying end of the sharpening (and later onsetting flattening; after passing $E_\text{int}=2.6n_0\hbar^2/m$ the curvature of the density at the minimum only increases slightly and soon decreases again) of the density allows the curvature of the interaction energy decay (graph in figure \ref{fig:simulationextrema}) to finally reach a minimum. Interestingly, according to (\ref{Eintlinear}), the interaction energy corresponds to a reciprocal Lorentz factor of $\sqrt{3/8}$. In other words, the minimum in curvature is reached and the acceleration of the decay in interaction energy hence maximal, when the analytical approximation (\ref{rhohighv}) is non-negative everywhere for the first time.

Afterwards, the decrease in interaction energy continues to accelerate, until the slope reaches a minimum at $E_\text{int}\approx2.0n_0\hbar^2/m$. Remarkably, this corresponds to a reciprocal Lorentz factor of $\epsilon\approx1/2=\sqrt{2/8}$. The soliton reaches a depth according to the analytical solution of $2n_0/3$, while the numerical solution shows a depth of $0.5n_0$. Finally, upon reaching $E_\text{int}\approx1.5n_0\hbar^2/m$, the curvature maximizes. Interestingly, this seems to coincide with the number damping induced maximum in the damping rate (see appendix \ref{Extrema}). Subsequently, the numerical damping rate agrees with our analytical result in the sign of the slope and the curvature, heralding a rapid increase of accuracy in our solution. This corresponds to a reciprocal Lorentz factor of $\sqrt{1/8}$ and an analytical depth of $n_0/3$, while the numerics suggest a depth of roughly $0.3n_0$.

We identified properties of the rapid decrease in interaction energy previous to the validity of our analytical result (\ref{energydamp}). The interaction energy under the influence of energy damping takes two extrema in curvature, and one in gradient. Interestingly, if we employ relation (\ref{Eintlinear}), these extrema appear for $\epsilon=\sqrt{1/8},\ \sqrt{2/8},\ \sqrt{3/8}$. This implies that until the first extremum in curvature is reached the depth of the analytical result (\ref{rhohighv}) takes negative values. Similarly, for the second extremum in curvature of the energy-damped interaction energy, the damping rate for number damping takes a minimum.

The argumentation for this section was based on a certain choice of $8n_0a_\text{s}^2N_\text{cut}$ and $l_z$. However, weak damping is linear in the former. As long as we do not leave the weak damping regime in varying $8n_0a_\text{s}^2N_\text{cut}$, the position of extrema is hence unaffected. Similar arguments imply that they are also robust with respect to changes in the thickness of the thermal cloud $l_z$, as long as $l_z$ does not become too small (see appendix \ref{Extrema}). Numerical simulations with different values of $l_z$ verify a negligible dependence.

The simulation also clearly demonstrates that after the annihilation of the vortex dipole the interaction energy is a useful quantity to study. As the interaction energy at the moment of annihilation is already considerably lower than in the PV regime, it might also shed light on the process of the death of a dipole.

\subsection{Pad{\'e} approximants near dipole annihilation}
\label{Pade}
Unfortunately, there is no analytical solution describing a JRS in the intermediate velocity regime around the annihilation. However, as argued before, we are interested in the interaction energy for which the behaviour close to the minimum gives the main contribution. For individual velocities the wave function can be approximated by a rational function of any given order, its Padé approximation of this order \cite{berloff_pade_2004}. We employ these approximations in the following to derive an approximation for the damping of the interaction energy close to the death of the dipole (\ref{Padedamp}).

Padé approximations of a JRS shortly before ($v\simeq 0.566c$), at ($v\simeq 0.636c$) and shortly after ($v=c/\sqrt{2}$) the annihilation are provided in \cite{berloff_pade_2004}. First, we calculate the interaction energies to be $2.93n_0\hbar^2/m$, $2.79n_0\hbar^2/m$ and $2.63n_0\hbar^2/m$, respectively. We derive via a linear fit an approximation for the dependence of the velocity on the interaction energy close to the annihilation
\begin{align}
	v(E_\text{int})&\simeq v_0-\nu\frac{mcE_\text{int}}{n_0\hbar^2},
\end{align}
where $v_0=1.943c$ and $\nu=0.469$. We use this approximation to express the damping of the velocity in the pELE through the interaction energy:
\begin{align}
		E_\text{int}\frac{d}{dt}\frac{1}{v}&=-\frac{E_\text{int}}{v^2}\frac{dv}{dt}\nonumber\\
		&\simeq\nu\frac{mcE_\text{int}}{n_0\hbar^2v^2}\frac{dE_\text{int}}{dt}.
\end{align}
Inserting these results into the pELE (\ref{edampi}) gives 
\begin{align}
		\frac{dE_\text{int}}{dt}\bigg|_\varepsilon&\simeq-\frac{n_0\hbar^2}{n_0\hbar^2v+\nu mcE_\text{int}}\frac{\hbar v^3}{2}\nonumber\\
		&\times\int \frac{d^2\textbf{k}}{(2\pi)^2}\tilde{\varepsilon}(\textbf{k})\left|\widetilde{\partial_xn}(\textbf{k})\right|^2.
\end{align}

The appearing integral can now be calculated using the Padé approximations. We should note that $\tilde{\varepsilon}$ depends on the thickness of the atomic cloud $l_z$. Numerical results are thus not general and we have to specify $l_z$. Here, we will calculate the integral for the parameters in experiment \cite{meyer_observation_2017}. However, as discussed in appendix \ref{Extrema}, we do expect an antiproportional dependence on $l_z$. Hence, the obtained results should be applicable in general. A linear fit of the resulting damping strength then gives an approximation for the damping rate valid around the annihilation
\begin{align}
	\label{Padedamp}
	\frac{dE_\text{int}}{dt}\bigg|_\varepsilon\simeq\frac{8n_0^2a_\text{s}^2N_\text{cut}\hbar^3}{m^2l_z\xi}\left(b_0+b_1\frac{mE_\text{int}}{n_0\hbar^2}\right),
\end{align}
where $b_0=-2.3$ and $b_1=0.7$. In contrast to the other damping rates found in this work, the damping of the interaction energy in the intermediate regime is not equivalent with the damping rate of the momentum. Nevertheless, this result provides an estimate for a directly measurable quantity in an experimentally less challenging regime than (\ref{energydamp}), which is only valid for shallow JRS with large extensions.

\begin{figure}
	\centering
	\includegraphics[width=1\linewidth]{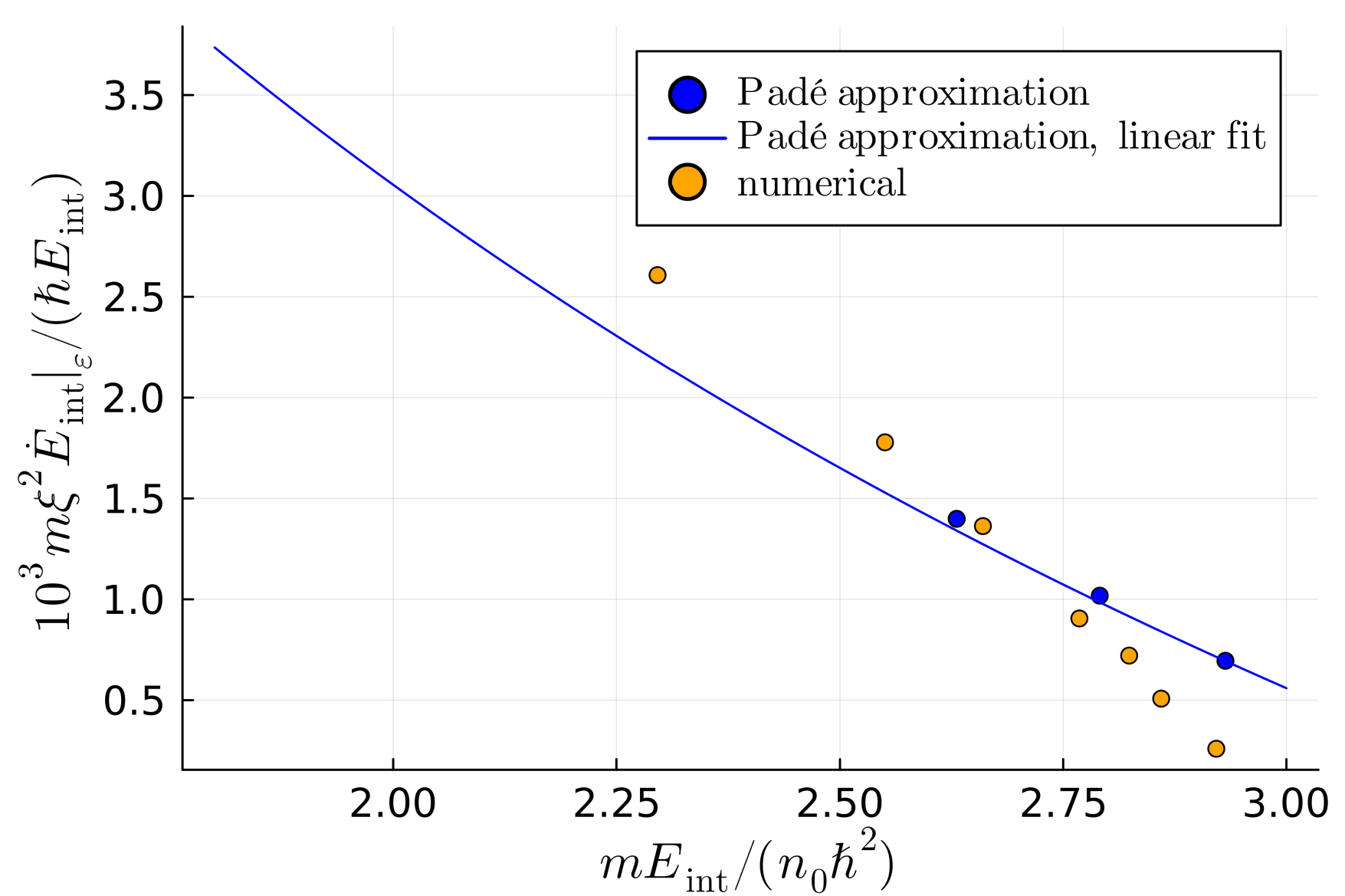}
	\caption{Normalized damping strength calculated using Padé approximations given in \cite{berloff_pade_2004} (blue dots), together with a (normalized) linear fit of the damping strength (blue line) and numerical results (orange dots). The Padé approximations give predictions in reasonable agreement with the numerics, especially in the regime after the annihilation (at $E_\text{int}\approx2.8n_0\hbar^2/m$, see section \ref{vortexannihilation}). The parameters used are again from experiment \cite{meyer_observation_2017}.}
	\label{fig:pade}
\end{figure}
Figure \ref{fig:pade} shows the obtained damping rates together with numerical results from the simulation. Good agreement with the predictions using the Padé approximations is demonstrated, especially in the regime after the annihilation. An improvement of the predictions in this regime might be due to the fact that only one minimum remains: the Padé approximations in the dipole regime needs to capture two minima, thereby restricting its ability to properly describe their form. After the annihilation, the approximation only needs to describe one minimum while being of same order. This is distinctly important in calculating the interaction energy, as this quantity depends mainly on the behaviour close to the minima. For improving the results found here, one thus would need to employ a Padé approximation of higher order in the dipole regime.

Our findings reveal an already strong damping of the interaction energy at and immediately after the annihilation. Hence, they demonstrate the usefulness of the interaction energy in characterising the properties of the JRS for all velocities but the low velocity regime of two widely separated vortices.
\section{Conclusions}\label{conc}
We investigated the thermally induced decay of planar JRS in the framework of the stochastic projected GPE, focusing on the damping mechanisms. Using Euler-Lagrange perturbation theory \cite{kivshar_lagrangian_1995}, we re-derived known results about the influence of energy damping \cite{mehdi_mutual_2023} in the low-velocity limit of two widely separated vortices. We found a logarithmic correction to the decay of a vortex dipole by number damping missing in previous work \cite{tornkvist_vortex_1997,mehdi_mutual_2023,sergeev_mutual_2023}; the correction is essential to capture number damping decay over large changes in dipole distance, and for tests of mutual friction~\cite{mehdi_mutual_2023} where $\gamma$ is not phenomenological. In the high-velocity rarefaction pulse regime we obtained analytical decay rates for both the number and energy damping processes, in good agreement with numerical simulations.

We derived bounds for the relative importance of the two damping processes in the high and low velocity regimes. Our key finding is that the dominant damping mechanism for typical planar BEC experimental parameters is the number-conserving energy transfer between atoms in the thermal cloud and the coherent region, referred to as energy damping. Only in a thin, hot BEC and for large scale excitations does this not hold; in that case the transfer of atoms between the coherent and incoherent regions accompanied by the growth of the condensate dominates the JRS decay. Energy damping is particularly strong in the intermediate regime, near the critical velocity for dipole collapse. In general the energy damping appears to be significant for excitations near the scale of the healing length.

The interaction energy provides a convenient measure of JRS decay in the high-velocity regime. Crucially, this quantity depends solely on the density and should be measurable in experiments, providing a direct test of JRS damping mechanisms (\ref{numberdamping}), (\ref{energydamp}) and thereby of SPGPE theory. Moreover, the rates for number (\ref{numberdamping}) and energy damping (\ref{energydamp}) are strongly dissimilar not only in their strength but also in their dependence on interaction energy. The pertinent damping process can be determined by considering the fractional rate of change of interaction energy: for number damping the rate is essentially constant, growing slightly with decreasing interaction energy, while energy damping instead weakens significantly. Testing this beahvior may be a promising route for experimental test of reservoir interaction theory. The interaction energy is also a  useful measure of decay in the intermediate velocity regime near the annihilation of the vortex dipole where it is strongly sensitive to the collapse, and the PV model loses its validity. As JRS in this regime are smaller and have more depth, the extraction of the interaction energy can be expected to be less challenging than in the high-velocity regime. In (\ref{Padedamp}) we provide a prediction of the energy damping rate close to the annihilation.

In future work it would be interesting to understand the influence of noise associated with each reservoir interaction mechanism on the JRS~\cite{gardiner_stochastic_2003,rooney_stochastic_2012,bradley_low-dimensional_2015}. While unimportant compared to the damping term at low temperatures (see appendix \ref{Noise}), thermal noise will be important for JRS decay in systems with finite temperature. It would be interesting to investigate the role of noise in the decay, the possibility of thermal excitations of rarefaction pulses and vortex dipoles, and their role in quantum turbulence. 
\section*{Acknowledgements}
We are grateful to Zain Mehdi for useful comments about the manuscript, and the Dodd-Walls Center for Photonic and Quantum Technologies for financial support.
\appendix
\section{Energy damping rate in the high-velocity limit}
\label{partrho}
We start by calculating $\widetilde{\partial_xn}(\textbf{k})$. It is
\begin{align}
	\begin{split}
		\widetilde{\partial_xn}(\textbf{k})&=-4\epsilon^2n_0ik_x\int d^2\textbf{r}e^{-i\textbf{k}\cdot\textbf{r}}\\
		&\times\frac{3/2+\epsilon^4y^2/\xi^2-\epsilon^2x^2/\xi^2}{(3/2+\epsilon^4y^2/\xi^2+\epsilon^2x^2/\xi^2)^2}\\
		&=-4\epsilon^2n_0ik_x\frac{\xi^2}{\epsilon^3}\frac{3}{2}\left(1-\frac{2\epsilon^4}{3\xi^2}\partial_{k_y}^2+\frac{2\epsilon^2}{3\xi^2}\partial_{k_x}^2\right)\\
		&\times\int d^2\tilde{\textbf{r}} e^{-i(\xi k_x\tilde{x}/\epsilon+\xi k_y\tilde{y}/\epsilon^2)}\frac{1}{(3/2+\tilde{y}^2+\tilde{x}^2)^2}.
	\end{split}\label{ftdxn}
\end{align}
We now calculate the integral. Switching to polar coordinates $r,\phi$ we see
\begin{align}
	\begin{split}
		&\int_0^\infty dr\frac{r}{(3/2+r^2)^2}\int_0^{2\pi}d\phi e^{i\sqrt{(\xi k_x/\epsilon)^2+(\xi k_y/\epsilon^2)^2}r\cos(\phi)}\\
		&=2\pi\int_0^\infty dr\frac{rJ_0(\sqrt{(\xi k_x/\epsilon)^2+(\xi k_y/\epsilon^2)^2}r)}{(3/2+r^2)^2}\\
		&=\pi\sqrt{\frac{2}{3}}\frac{\xi}{\epsilon}\sqrt{k_x^2+\frac{k_y^2}{\epsilon^2}}K_1\left(\sqrt{\frac{3}{2}}\frac{\xi}{\epsilon}\sqrt{k_x^2+\frac{k_y^2}{\epsilon^2}}\right).
	\end{split}
\end{align}
Noting that the derivative terms cancel out in \eref{ftdxn} due to their respective $\epsilon$ scaling and the symmetry of the integrand, we conclude
\begin{align}
	\widetilde{\partial_xn}(\textbf{k})&=8\pi n_0i\sqrt{\frac{2}{3}}\xi\kappa^2\cos^3(\varphi)K_1(\kappa)
\end{align}
with the $k$-space elliptical coordinates given by  $\kappa=\sqrt{3/2}\xi/\epsilon(k_x^2+k_y^2/\epsilon^2)^{1/2}$ and $\kappa\cos(\varphi)=\sqrt{3/2}(\xi/\epsilon) k_x$ and $K_1$ the modified Bessel-function of second kind of order one.

Next we calculate the integral in (\ref{edampi}). We transform to the momentum space variables $\kappa,\ \varphi$. The Bessel-function has an asymptotic expansion for small arguments
\begin{align}
	K_0(z)&\sim-\ln\left(\frac{z}{2}\right)-\gamma_\text{Euler},
\end{align}
with $\gamma_\text{Euler}=0.577...$ the Euler-Mascheroni constant. Hence, we can approximate the kernel $\tilde{\varepsilon}$ in the integral by (most of our arguments require $\epsilon\ll1$; however, this step requires the usually slightly stronger condition $\epsilon\ll\xi/l_z$)
\begin{align}
	\begin{split}
		&\tilde{\varepsilon}\left(\sqrt{\frac{2}{3}}\frac{\epsilon}{\xi} \kappa\cos(\varphi),\sqrt{\frac{2}{3}}\frac{\epsilon^2}{\xi} \kappa\sin(\varphi)\right)\\
		\sim&8N_\text{cut}a_\text{s}^2\left[-\ln\left\{\frac{1}{12}\kappa^2\cos^2(\varphi)\left(\frac{\epsilon l_z}{\xi}\right)^2\right\}-\gamma_\text{Euler}\right].
	\end{split}
\end{align}
We obtain
\begin{align}
	\begin{split}
		\frac{dE_\text{int}}{dt}\bigg|_\varepsilon&=\frac{128N_\text{cut}a_\text{s}^2n_0^2\hbar^3}{9m^2\xi^2}\epsilon^3\int_0^\infty d\kappa\kappa^5K_1(\kappa)^2\int_0^{2\pi}d\varphi\\
		&\times\cos^6(\varphi)\bigg[-\ln\left(12\right)+\gamma_\text{Euler}+2 \ln\left(\frac{\epsilon l_z}{\xi}\right)\\
		&+2\ln(\kappa)+\ln\{\cos^2(\varphi)\}\bigg].
	\end{split}
\end{align}
The appearing integrals give
\begin{align}
	\begin{split}
		\int_0^{2\pi}d\varphi \cos^6(\varphi)=&\frac{5\pi}{8},\\
		\int_0^{2\pi}d\varphi \cos^6(\varphi)\ln\{\cos^2(\varphi)\}=&-\frac{5\pi}{4}\ln(2)+\frac{37\pi}{48},\\
		\int_0^\infty d\kappa\kappa^5K_1(\kappa)^2=&\frac{8}{5},\\
		\int_0^\infty d\kappa\kappa^5K_1(\kappa)^2\ln(\kappa)=&\frac{76}{75}-\frac{8}{5}\gamma_\text{Euler}+\frac{4}{5}\ln(4),
	\end{split}
\end{align}
which together with (\ref{Eintlinear}) then gives the result (\ref{energydamp}) in the main text.

\section{Validity of the Lagrangian Method and Stability of JRS}
\label{Validity}
To derive the damping rates in this work, we used the perturbed Euler-Lagrange equation (\ref{pELE}). Their use is restricted to the condition of weak damping. However, it is not readily obvious what ``weak damping'' means in this context. Certainly, it should be compared to the unitary time evolution. A key assumption is that we preserve a JRS at all times. In the unitary time evolution, the velocity with which information travels is the speed of sound $c$. The damping is not uniformly distributed over the extension of the JRS. To allow the wavefunction to contain the form of a JRS thus requires the information about a change in its form to be transported over its entire extension faster than the change occurs. We want to analyze in this section, under which condition this holds in the two regimes of slow and fast solitons.

Let us start with slow JRS. We hence look at two widely separated vortices. The parameter that governs the form of the soliton in this regime is $v/c$. For its change we have for energy damping using (\ref{MehdiResult})
\begin{align}
	\begin{split}
		\frac{d}{dt}\frac{v}{c}&=-\sqrt{2}\frac{\xi}{d_\text{v}}\frac{\dot{d_\text{v}}}{d_\text{v}}=4\sqrt{2}\xi n_0a_\text{s}^2N_\text{cut}F\left(\frac{l_z^2}{8\xi^2}\right)\frac{\hbar}{m}\frac{1}{d_\text{v}^3}
	\end{split}
\end{align}
and for number damping using (\ref{numberd})
\begin{align}
	\frac{d}{dt}\frac{v}{c}&=2\sqrt{2}\gamma \xi\frac{\hbar}{m}\ln\left(\alpha\frac{d_\text{v}}{\xi}\right)\frac{1}{d_\text{v}^3}.
\end{align}
The extension of the JRS is characterized by the distance between the two vortices $d_\text{v}$. Hence, the rate with which information about the form of the dipole is transported over its entire extension is given by $c/d_\text{v}$. This needs to be large against the rate of the change
\begin{align}
	\frac{c}{d_\text{v}}&\gg\frac{d}{dt}\frac{v}{c}.
\end{align}
We  end up with
\begin{align}
	1\gg&8n_0a_\text{s}^2N_\text{cut}F\left(\frac{l_z^2}{8\xi^2}\right)\frac{\xi^2}{d_\text{v}^2},
\end{align}
for energy and
\begin{align}
	1\gg&4\gamma\ln\left(\alpha\frac{d_\text{v}}{\xi}\right)\frac{\xi^2}{d_\text{v}^2},
\end{align} 
for number damping. These results suggest that the form of a JRS in this regime will usually be preserved. Only for such dense condensates or long-reaching interactions, that thermal atoms can not pass the condensate without a scattering event, might energy damping trigger the loss of the solitonic form. 

We now consider the high-velocity regime. The extension of a fast moving soliton is proportional to $\xi/\epsilon^2$. Hence, the rate with which information about the form is transmitted over the entire JRS is proportional to $c/(\xi/\epsilon^2)$. The relevant parameter that describes the form is the reciprocal Lorentz factor $\epsilon$. Thus, the condition
\begin{align}
	\frac{c}{\xi/\epsilon^2}\gg \left|\frac{d\epsilon}{dt}\right|
\end{align}
needs to be fulfilled. In the case of energy damping we obtain using (\ref{epsilonenergydamp})
\begin{align}
	\label{EnergyInstability}
	1\gg -\frac{32\sqrt{2}}{3}n_0a_\text{s}^2N_\text{cut}\epsilon\left[2\ln\left(\frac{\epsilon l_z}{\xi}\right)+C_E\right].
\end{align}
In the case of number damping we obtain using (\ref{numberdamp})
\begin{align}
	\label{NumberInstability}
	1\gg \sqrt{2}\gamma \frac{1}{\epsilon}.
\end{align}
Hence, while weak energy damping in principle allows a rarefaction pulse to stay such for an indefinit time, under number damping it is inherently unstable and will eventually dissolve after a finit time.

In the simulations the instability is noticeable in that approaching the bounds (\ref{EnergyInstability}) and (\ref{NumberInstability}) the soliton tends to become curved.

\section{Simulation procedure}
\label{Simulation}
\begin{figure}[tb]
	\centering
	\includegraphics[width=1\linewidth]{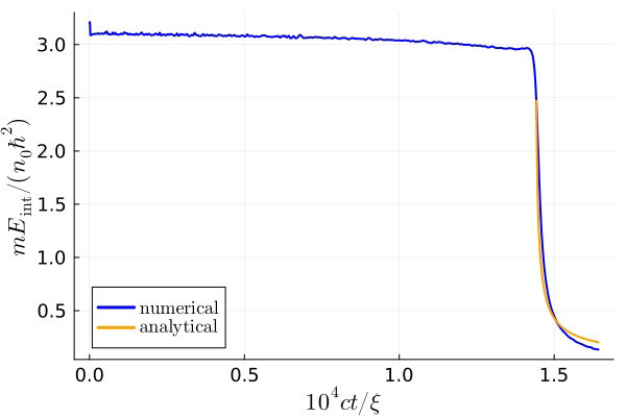}
	\caption{Interaction energy calculated from the numerical solution of the $\varepsilon$GPE with $8a_\text{s}^2n_0N_\text{cut}=10^{-2}$, $l_z=\sqrt{2}\xi$ (blue line). The simulation starts with two vortices separated by $12\sqrt{2}\xi$. As long as there are two clearly identifiable vortices, the interaction energy shrinks only slightly. After roughly $1.4\times 10^4\xi/c$ the two vortices collapse and form a rarefaction pulse, resulting in a sudden drop of the interaction energy. Our analytical result (\ref{edampi}) captures the decline in interaction energy well (orange line). For the chosen parameters, energy damping is not sufficiently small compared to the unitary evolution in the high velocity limit. This results in the numerical solution deviating from a JRS, which in turn strengthens the damping of the interaction energy, leading to the discrepancy between analytics and numerics for long times. The drop of interaction energy at the beginning results from the initial condition being no exact solution of the GPE. This also leads to sound being created at the edges of the domain, which is responsible for the wiggles prominently observable in the dipole regime.}
	\label{fig:simulation}
\end{figure}
To obtain reliable initial conditions in the absence of a known analytical solution for JRS at intermediate velocities, one can either solve the time independent GPE or start with two vortices and let them collapse. Towards the later aim, we run a simulation of the $\varepsilon$GPE starting with two vortices at a distance of $12\sqrt{2}\xi$ and energy damping of strength $8a_\text{s}^2n_0N_\text{cut}=10^{-2}$, $l_z=\sqrt{2}\xi$ in the absence of number damping.

Figure \ref{fig:simulation} shows the full simulation. It demonstrates an only slight change in interaction energy prior to the collapse of the two vortices into a rarefaction pulse. As soon as the collpase occurs, we observe a sudden drop in the interaction energy. Thus, the interaction energy is shown to be a good candidate for the identification of the collapse and the analysis of JRS. Remarkably, our analytical result (\ref{edampi}) shows good agreement to the numerics. This is despite the energy damping used being not weak and although the JRS is observed to clearly bend over time.

As noted, energy damping becomes large compared to the unitary evolution for JRS in the high velocity regime. Hence, if we start a simulation with a rarefaction pulse, it loses its form. Therefore, we run a second simulation for use in the main text with a tenth of the damping strength $8a_\text{s}^2n_0N_\text{cut}=10^{-3}$. We started with a wavefunction $\psi$ derived from the first simulation close to the disappearance of two minima. From this latter simulation, all initial conditions in this work are derived.

\section{System Parameters}
\label{expparameters}
To understand the role of energy damping in experiments, we performed simulations and calculated the analytical damping rates using parameters given in \cite{meyer_observation_2017}. However, the parameters given are for the 3D GPE, while our arguments are valid in 2D. To derive the parameters relevant for the two dimensional dynamics, we assumed that the coherent part is in the ground state along the $z$-axis (the axis with a much stronger confinement than the other two axes). Integrating along the $z$-axis yields the following parameters: $l_z=\sqrt{\hbar/m\omega_z}$
for the extension of the gas along the $z$-axis, where $\omega_z$ is the trapping frequency along this axis,
$\mu=\mu_\text{3D}- \hbar\omega_z/2$, and $		\xi=\hbar/(\sqrt{2m\mu})$
for the 2D chemical potential and the healing length respectively (where $\mu_{3D}$ corresponds to the 3D chemical potential). The interaction strength $	g=\sqrt{8\pi}\hbar^2a_\text{s}/(ml_z)$ relates the background density $n_0$ and chemical potential via $\mu=gn_0$, which we use throughout as the two dimensional density.

The validity of the dimensional reduction was analyzed in \cite{salasnich_effective_2002}. It is suggesetd that a $x,y$ dependent choice of the thickness varying essentially between $l_z$ and roughly $1.2l_z$ would be appropriate for the experimental parameters. Hence, for an order of magnitude estimate we assume the system is in the $z$-direction ground state.

No temperature is reported in \cite{meyer_observation_2017}; unless otherwise stated our results are calculated for a temperature of 30nK. Moreover, a choice of a proper cutoff $\epsilon_\text{cut}$ is crucial for the validity of the $\varepsilon$GPE and $\gamma$GPE. It should be choosen so that the number of particles at the cutoff $N_\text{cut}$ is of order one. Therefore, we choose throughout a cutoff leading to $N_\text{cut}=1$.

In figure \ref{fig:comparison} we mark the critical temperature $T_\text{c}$. We estimated it using the result for an ideal Bose gas in a harmonic trap \cite{dalfovo_theory_1999}
\begin{align}
k_\text{B}T_\text{c}&=\hbar\left(\frac{\omega_x\omega_y\omega_zN}{\zeta(3)}\right)^{1/3}.
\end{align}
Here, $\omega_j$ is the trapping frequency along the $j$-axis ($j=x,y,z$) and $\zeta$ denotes the Riemann-zeta function, while $N$ is the total particle number~\footnote{Usually the finite size correction due to the ground state energy is significant for low dimensional systems and would be required for complete quantitative modelling of an experiment.}.

To make quantitative predictions, a more careful dimensional reduction may be necessary. Furthermore, we did not take great care in choosing the optimal cutoff for the given experiment. However, in this paper we mainly aim on establishing the validity of our approach and determining the importance of energy damping in the regime of feasible experiments.

\section{Correction of Number Damping Simulations}
\label{simulationcorrection}
\begin{figure}
	\centering
	\includegraphics[width=1\linewidth]{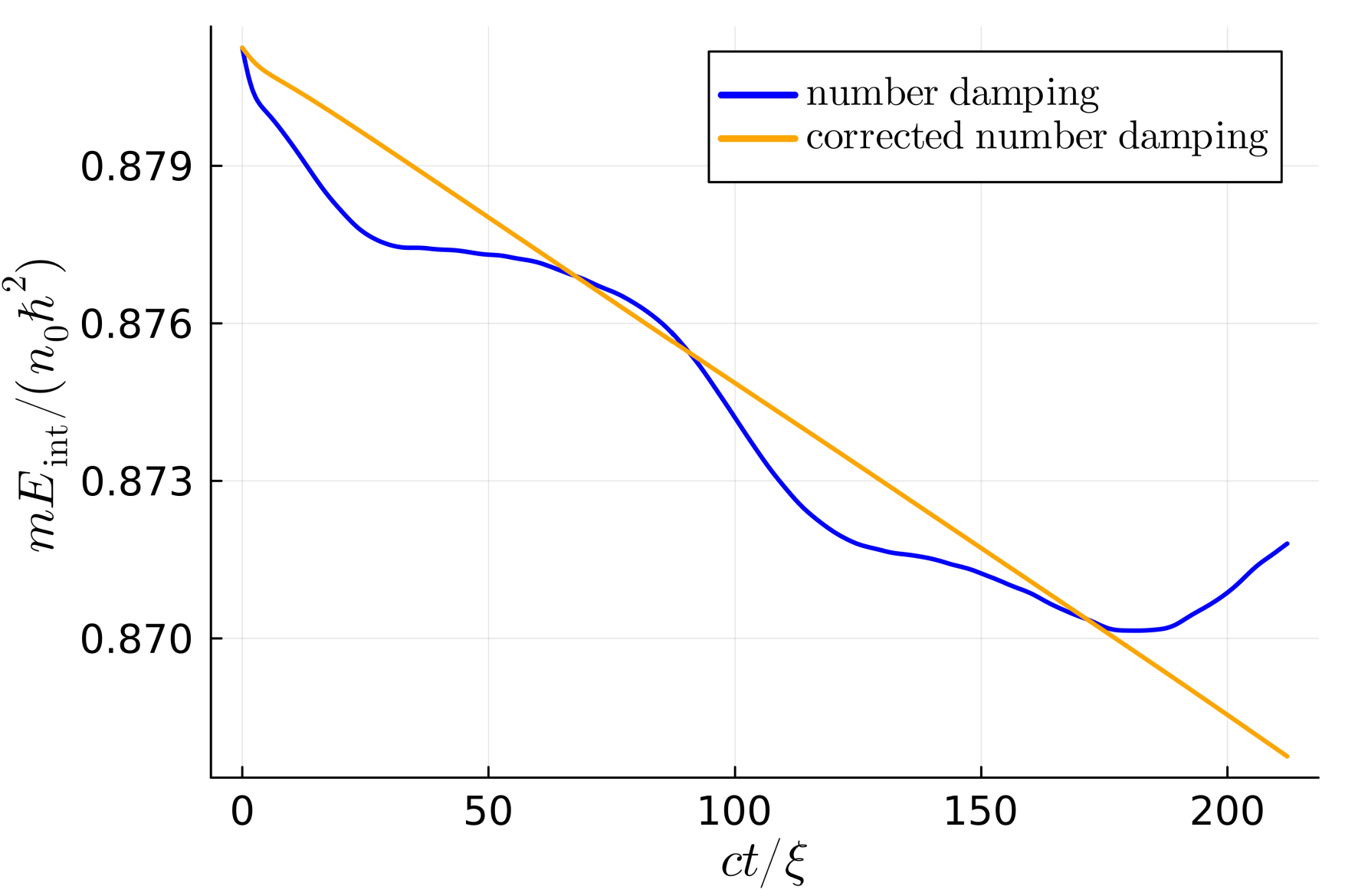}
	\caption{Interaction energy from a simulation of the $\gamma$GPE (blue) and from the same simulation subtracted by a simulation of the undamped GPE (orange). The subtraction of the undamped result eliminates the sound but keeps the overall decrease in interaction energy due to damping.}
	\label{fig:correctednumber}
\end{figure}
For the experimental values in \cite{meyer_observation_2017} used in this work to analyze the damping strength, we observed number damping to be much weaker than energy damping. Indeed, it turns out that the effect of number damping is not considerably stronger than the influence of the sound present in the simulations due to non exact initial conditions. Although the damping process turns out to be quite robust to the influence of sound over the long term, on short timescales this leads to difficulties in extracting the damping rate numerically. Taking long time evolutions for computation of the damping is no solution, since the damping rates change with time. To solve these difficulties, we run two simulations starting at the same initial conditions: one with ($\gamma$GPE), and one without (GPE) number damping. The latter one should contain the information about sound present, which is not strongly affected by number damping in the short term. Despite this, it does not contain any additional damping. Hence, we should be able to obtain a corrected result for the strength of number damping by subtracting the interaction energies obtained at the presence of number damping by the undamped ones. Figure \ref{fig:correctednumber} shows that the effects of sound are indeed essentially eliminated, while the overall damping is not affected.

Energy damping is much stronger than number damping and the influence of sound therefore less pronounced. However, for the sake of consistency, we again subtracted the damping free GPE case in order to calculate the damping rates.

\section{Higher Order Corrections}
\label{higherordercorrections}
Our main results (\ref{numberdamping}) for number and (\ref{energydamp}) for energy damping were derived in the lowest order of the reciprocal Lorentz factor $\epsilon$. In this section, we have a look on higher-order corrections. For number damping, the perturbed Euler-Lagrange equation (\ref{pELE}) leads to
\begin{align}
	\frac{d}{dt}\frac{E_\text{int}}{v}\bigg|_\gamma=-\frac{\gamma mv}{\hbar}E_\text{tot}
\end{align}
In the case of energy damping, we have
\begin{align}
	\frac{d}{dt}\frac{E_\text{int}}{v}\bigg|_\varepsilon=&-\frac{\hbar v}{2}\int \frac{d^2\textbf{k}}{(2\pi)^2}\left|\widetilde{\partial_xn}(\textbf{k})\right|^2\tilde{\varepsilon}(\textbf{k}).
\end{align}

So far, the only assumption we made is that the damping is weak and $\psi$ thus in the form of a JRS at all times. In the main text, we approximated the velocity $v$ in the pELE by the speed of sound $c$. We note $v^2=c^2(1-\epsilon^2)$ and
\begin{align}
	\begin{split}
		E_\text{int}\frac{d}{dt}\frac{1}{v}&=-E_\text{int}\frac{dv/dt}{v^2}=\frac{E_\text{int}}{v}\frac{\epsilon}{1-\epsilon^2}\frac{d\epsilon}{dt}\\
		&\simeq\frac{1}{v}\left(\frac{4\pi}{3}\frac{n_0\hbar^2}{m}\right)^{-2}E_\text{int}^2\frac{dE_\text{int}}{dt},
	\end{split}
\end{align}
where the last line is valid in lowest order of $\epsilon$ (see (\ref{Eintlinear})). With this, we conclude
\begin{align}
	\frac{dE_\text{int}}{dt}\bigg|_\gamma=-\frac{\gamma \hbar}{2m\xi^2}E_\text{tot}\left[1-2\left(\frac{4\pi}{3}\frac{n_0\hbar^2}{m}\right)^{-2}E_\text{int}^2\right]
\end{align}
and
\begin{align}
	\label{higherorder}
	\begin{split}
		\frac{dE_\text{int}}{dt}\bigg|_\varepsilon=&-\frac{\hbar^3}{4m^2\xi^2}\int \frac{d^2\textbf{k}}{(2\pi)^2}\left|\widetilde{\partial_xn}(\textbf{k})\right|^2\tilde{\varepsilon}(\textbf{k})\\
		&\times\left[1-2\left(\frac{4\pi}{3}\frac{n_0\hbar^2}{m}\right)^{-2}E_\text{int}^2\right]
	\end{split}
\end{align}
for number and energy damping, respectively. Both are valid two orders in $\epsilon$ and thus $E_\text{int}$ higher then the results in the main text.

\begin{figure}[tb]
	\centering
	\includegraphics[width=1\linewidth]{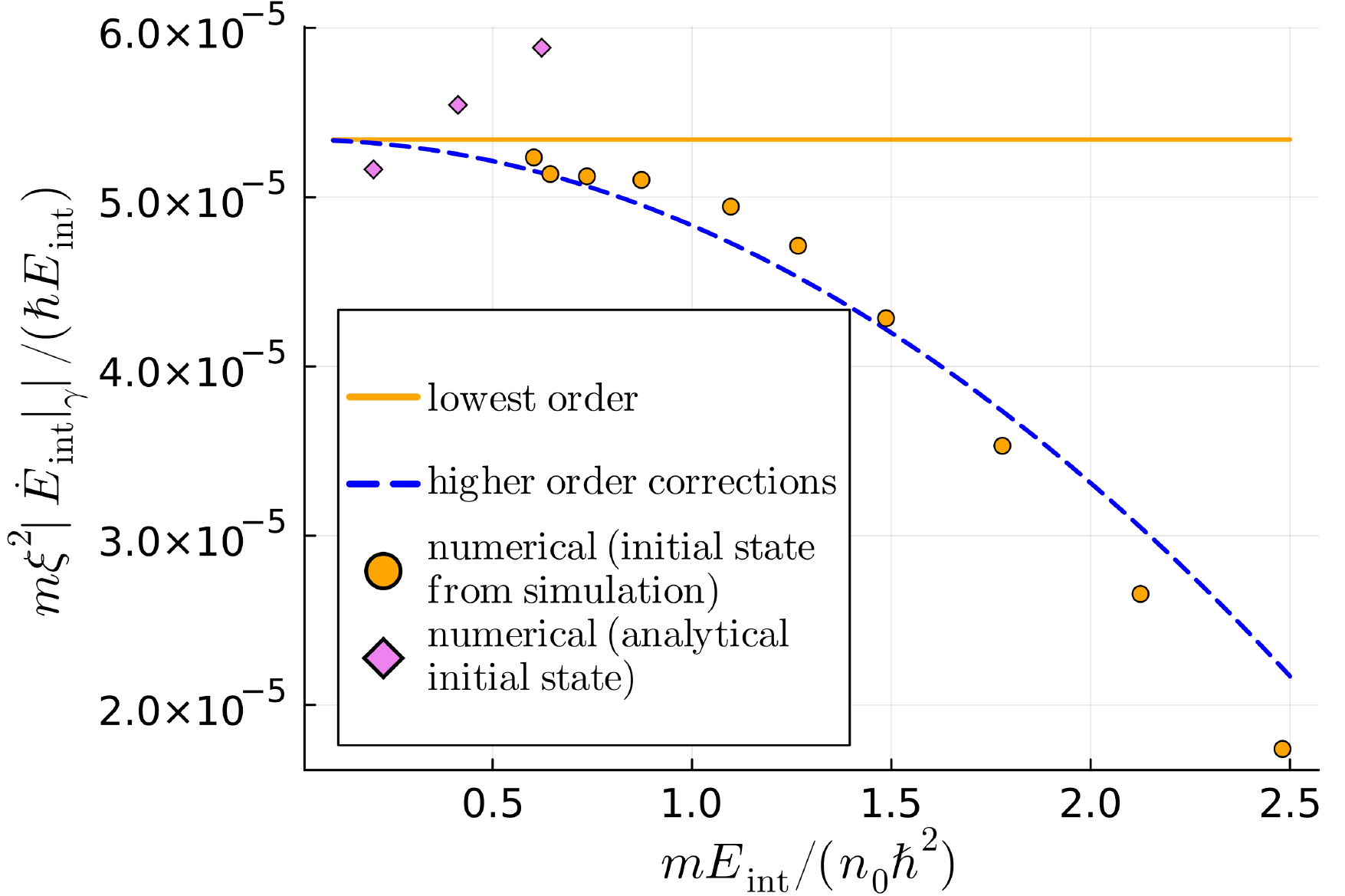}
	\caption{Same graph as (a) in \fref{fig:dEintnumber}, but including corrected results for the damping rates (blue dashed line). The higher order corrections found in (\ref{higherordernumber}) already improve drastically on the lowest order predictions of the number damping rate at intermediate interaction energies.}
	\label{fig:deintnumberhigherorder}
\end{figure}
For number damping, we went on approximating $\psi$ by its lowest order expansion in the reciprocal Lorentz factor $\epsilon$. We can calculate the next higher order in the total energy $E_\text{tot}$ without computing higher orders in $\psi$. According to (\ref{Integral-relation1}) it is
\begin{align}
	E_\text{tot}=2E_\text{int}+\frac{\hbar^2}{m}\int d^2\textbf{r}|\partial_y \psi|^2.
\end{align} 
Direct calcultaion of the integral in lowest order of the reciprocal Lorentz factor $\epsilon$ using (\ref{Eintlinear}) and (\ref{theta}) yields $2/3\epsilon^2E_\text{int}$. Hence, we end up with
\begin{align}
	E_\text{tot}=2E_\text{int}+\frac{2}{3}\epsilon^2E_\text{int}.
\end{align}
Therefore, the damping is up to terms of at least the order $E_\text{int}^4$ given by
\begin{align}
	\label{higherordernumber}
	\frac{dE_\text{int}}{dt}\bigg|_\gamma=-\frac{\gamma \hbar}{m\xi^2}E_\text{int}\left[1-\frac{5}{3}\left(\frac{4\pi}{3}\frac{n_0\hbar^2}{m}\right)^{-2}E_\text{int}^2\right].
\end{align}
Figure \ref{fig:deintnumberhigherorder} demonstrates that the next higher order found in this section improves significantly on our findings in the main text.

\begin{figure}[tb]
	\centering
	\includegraphics[width=1\linewidth]{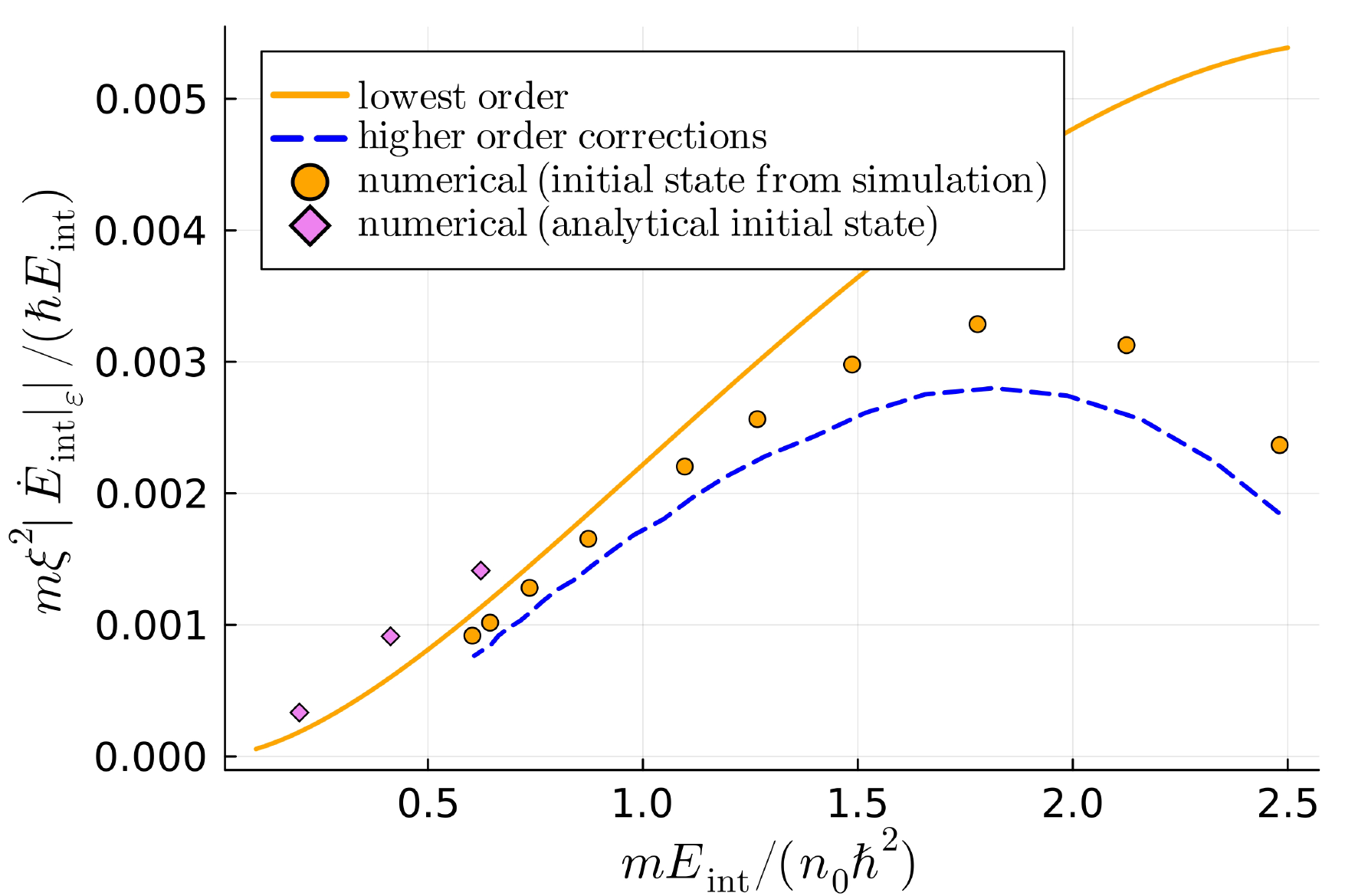}
	\caption{Same graph as (b) in \fref{fig:dEintnumber}, but including corrected results for the damping rates (blue dashed line). The higher order corrections correspond to the result found in (\ref{higherorder}), where the integral is calculated numerically from the initial conditions. For intermediate interaction energies, we observe a clear improvement compared to the lowest order.}
	\label{fig:deinthigherorder}
\end{figure}
In the analysis of energy damping, we again went on approximating $\psi$ and hence $n$ by its lowest order expansion in the reciprocal Lorentz factor $\epsilon$. We do not calculate the next higher order for $n$ here, although as in the number damping case we expect it to be on the same order as the corrections taken into account. Further, in the main text, we assume that $\tilde{\varepsilon}$ can be approximated in lowest order. Next higher orders will again be of comparable magnitude. Additionally, the argument of $\tilde{\varepsilon}$ depends on the reciprocal Lorentz factor $\epsilon$, which is only linear in the interaction energy in lowest order. Hence, this will lead to corrections on the same order, too. Instead of tackling these challenges analytically, we computed the integral appearing in (\ref{higherorder}) numerically from the initial conditions. As can be seen in figure \ref{fig:deinthigherorder}, the inclusion of the higher order leads to a considerable improvement at intermediate interaction energies. We should remark that the integral has a purely positive integrand and is calculated over a finite grid. Hence, the received value will in general be to small. For larger interaction energies this should not play a significant role, as the extension of the JRS is still well in the limits of the grid. For small interaction energies, however, the JRS in the simulation already reaches an extension that is barely restricted to the grid limits. This might explain the discrepancy at small interaction energies.

\section{Number Damping Bounds}
\label{numberstrength}
In the main text we analyzed conditions for energy damping to be the dominant mechanism in the decay of vortex dipoles and rarefaction pulses. Here we derive bounds for the dimensionless number damping strength~\cite{bradley_bose-einstein_2008} given in (\ref{gamdef}).
To simplify the sum we use the Lerch transcendent definition $\Phi[z,s,\alpha]=\sum_{n=0}^\infty z^n/(n+\alpha)^s$ 
to write it as
\begin{align}
	\sum_{j=1}^\infty z^j\Phi\left[z,1,j\right]^2&=\sum_{j=1}^\infty z^j\sum_{n,m=0}^\infty \frac{z^{n+m}}{(n+j)(m+j)}\nonumber\\
		&=\sum_{n,m=0}^\infty \sqrt{z}^{n+m}\sum_{j=1}^\infty \frac{\sqrt{z}^{n+j}}{n+j}\frac{\sqrt{z}^{m+j}}{m+j}.
\end{align}
We now observe that the summand on the right can be written as two integrals and obtain
\begin{align}
		&\sum_{n,m=0}^\infty \sqrt{z}^{n+m}\int_0^{\sqrt{z}}dx\int_0^{\sqrt{z}}dy\sum_{j=1}^\infty x^{n+j-1}y^{m+j-1}\nonumber\\
		&=\sum_{n,m=0}^\infty \sqrt{z}^{n+m}\int_0^{\sqrt{z}}dx\int_0^{\sqrt{z}}dyx^ny^m\frac{1}{1-xy}\nonumber\\
		=&\int_0^{\sqrt{z}}dx\int_0^{\sqrt{z}}dy\frac{1}{1-\sqrt{z}y}\frac{1}{1-\sqrt{z}x}\frac{1}{1-xy}\nonumber\\	\label{G5}
		&=\int_0^1dy\ln\left(\frac{1-zy}{1-z}\right)\frac{1}{(1-y)(1-zy)},
\end{align} 
where we used the geometric series (note that $\mu_{3D}<2\epsilon_\text{cut}$ and hence $\sqrt{z},x,y<1$), and
integrated over $x$ and performed a transformation $y\rightarrow y/\sqrt{z}$.

The integrand is monotonically increasing in the integration interval. An upper bound for the integral is therefore given by taking for the integrand its value at the upper integration limit
\begin{align}
	\frac{1}{1-z}\lim\limits_{y\nearrow1}\frac{\ln\left(\frac{1-zy}{1-z}\right)}{1-y}=\frac{z}{(1-z)^2}.
\end{align}
Together with the factor $e^{\beta\mu_{3D}}$, we find the upper bound
\begin{align}
		&e^{\beta\mu_{3D}}\sum_{j=1}^\infty z^j\Phi\left[z,1,j\right]^2\leq 
		\frac{e^{2\beta(\mu_{3D}-\epsilon_\text{cut})}}{(1-e^{\beta(\mu_{3D}-2\epsilon_\text{cut})})^2}\nonumber\\
		&=\frac{1}{(e^{\beta(\epsilon_\text{cut}-\mu_{3D})}-e^{-\beta\epsilon_\text{cut}})^2}
		\leq N_\text{cut}^2.
\end{align}

On the other hand, we can get a lower bound by approximating the integrand by its value at the lower integration limit
\begin{align}
	\begin{split}
		\ln\left(\frac{1}{1-z}\right)=&\ln\left(1+\frac{z}{1-z}\right)
		\geq z.
	\end{split}
\end{align}
We then find
\begin{align}
	e^{\beta\mu_{3D}}\sum_{j=1}^\infty z^j\Phi\left[z,1,j\right]^2\geq e^{2\beta(\mu_{3D}-\epsilon_\text{cut})}=\frac{N_\text{cut}^2}{(N_\text{cut}+1)^2}
\end{align}
as a lower bound. We thus have the useful bounds on the SPGPE dimensionless number damping 
\begin{align}\label{gammabounds}
	\frac{N_\text{cut}^2}{(N_\text{cut}+1)^2}<\frac{\gamma \lambda_\text{th}^2}{8a_\text{s}^2} <N_\text{cut}^2.
\end{align}


\section{Time of Annihilation and Extrema of the Damped Interaction Energy}
\label{Extrema}
\begin{figure}
	\centering
	\includegraphics[width=1\linewidth]{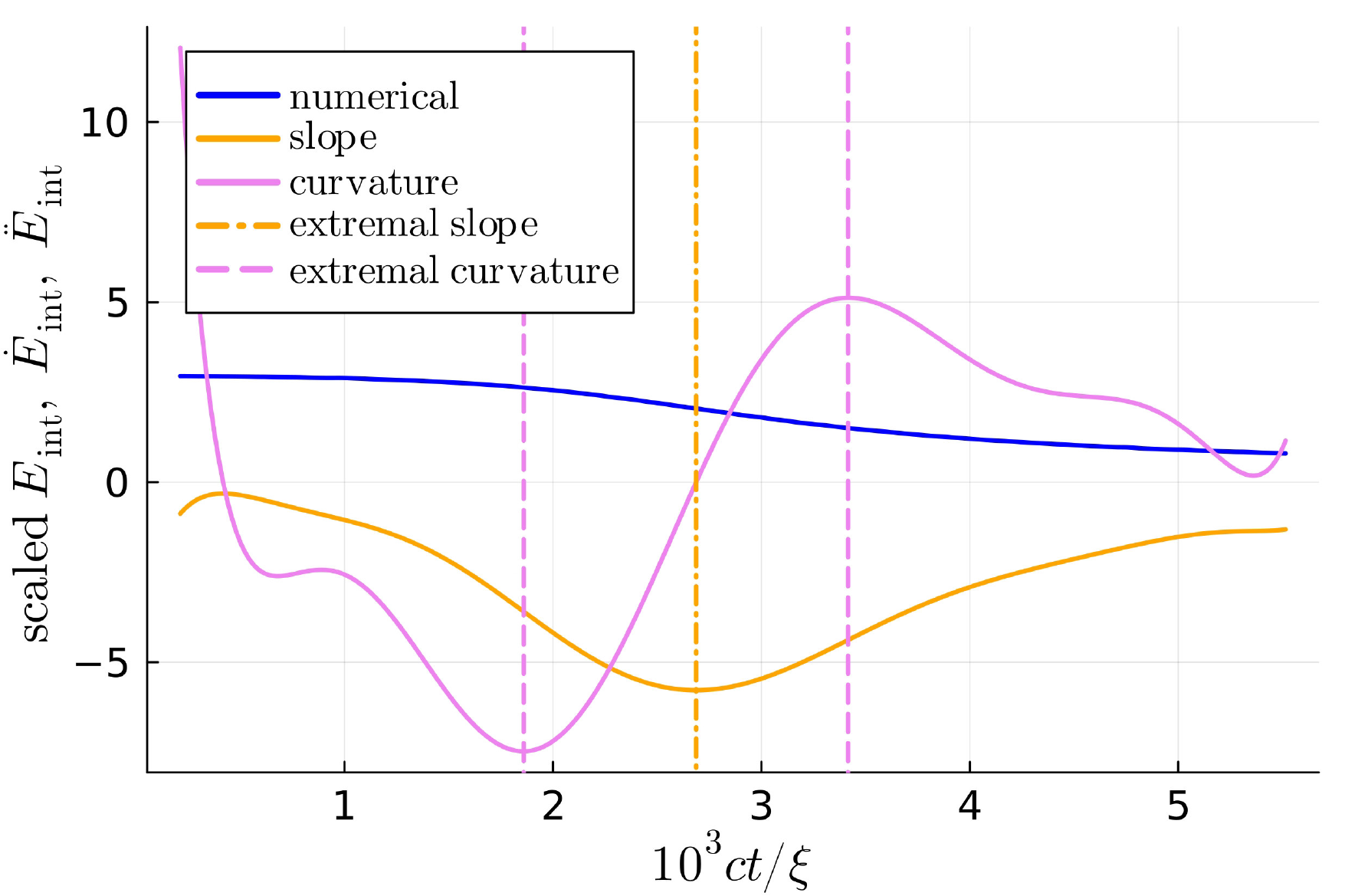}
	\caption{Interaction energy calculated from a simulation with $8a_\text{s}^2n_0N_\text{cut}=10^{-3}$, $l_z=\sqrt{2}\xi$ together with derivatives of a polynomial fit. These derivatives were used to determine the extrema of the slope (orange dash-dotted line) and curvature (violet dashed line) discussed in section \ref{vortexannihilation}.}
	\label{fig:extremabestimmen}
\end{figure}
\begin{figure}
	\centering
	\includegraphics[width=1\linewidth]{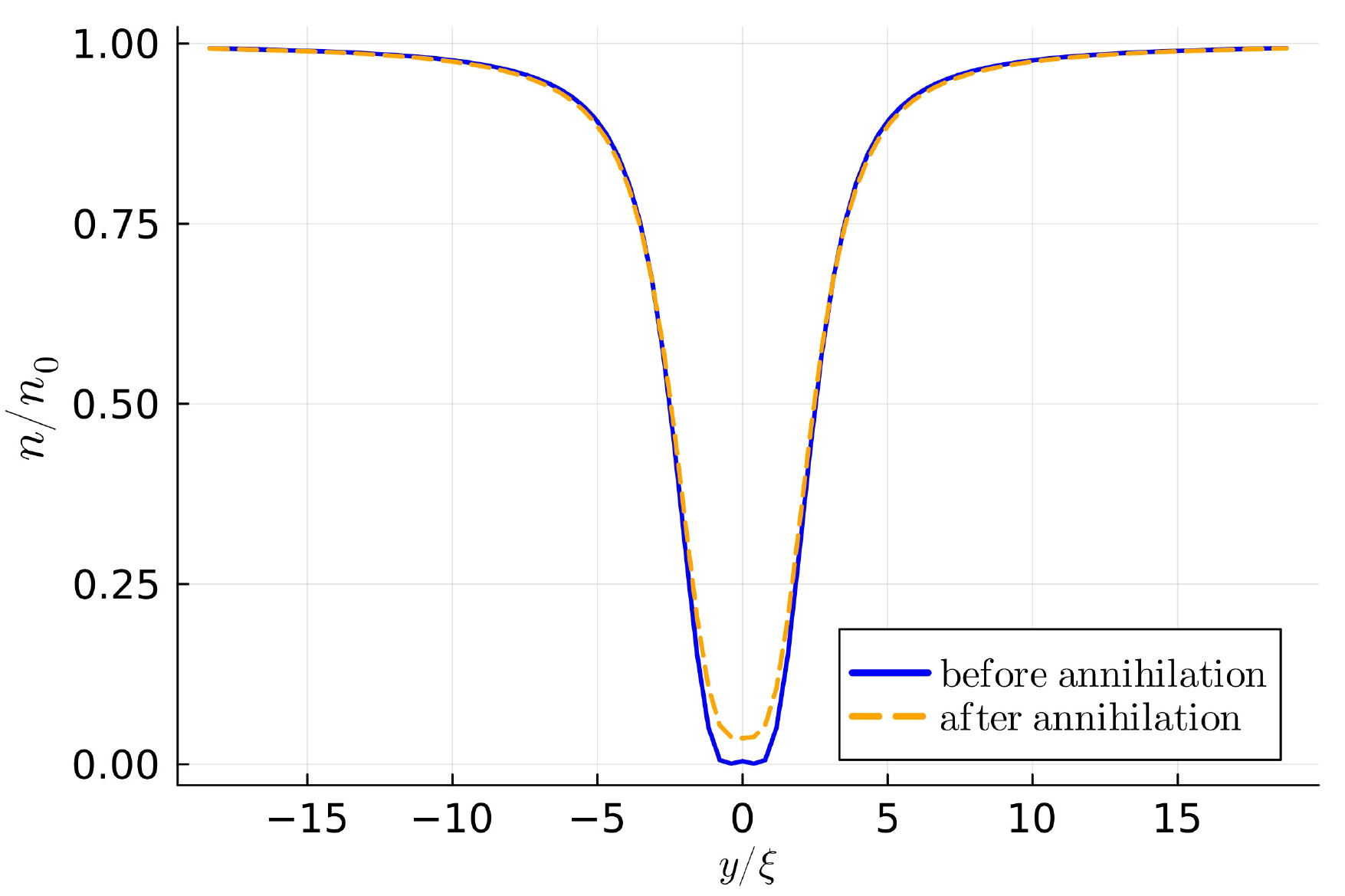}
	\caption{Profile of the density at the minimum along the y-axis shortly before (blue, solid line) and after (orange, dashed line) the annihilation occurs. The two minima still visible in the blue line reduce to one and do not reach zero anymore in the orange one. We deduce that the collapse has to occur between the times of the two profiles shown.}
	\label{fig:vornachkollaps}
\end{figure}

In section \ref{vortexannihilation} we discussed the moment of annihilation and the appearing extrema in the slope and curvature of the interaction energy. The annihilation was identified by the disappearance of two minima (see figure \ref{fig:vornachkollaps}). We derived the values of the extrema by a polynomial interpolation of the simulation data shown in figure \ref{fig:extremabestimmen}.

Our argumentation in the main part is based on a certain choice for the thickness of the atomic cloud $l_z$. Here, we justify the derivation of general statements from numerical observations of a particular choice. According to (\ref{edampi}) the damping rate is given by
\begin{align}
	\begin{split}
		\frac{dE_\text{int}}{dt}\bigg|_\varepsilon=&\frac{-\hbar v^2}{2(1-E_\text{int}[\partial_{E\text{int}}v]/v)}\notag\\
  &\times\int \frac{d^2\textbf{k}}{(2\pi)^2}\tilde{\varepsilon}(\textbf{k})\left|\widetilde{\partial_xn}(\textbf{k})\right|^2.\\
	\end{split}
\end{align}
The prefactor on the right hand side is independent of $l_z$. We hence only need to consider the integral. We will argue that we can use the asymptotic (large argument) expansion of $\tilde{\varepsilon}$. We are interested in the regime after the annihilation but before the validity of the analytical approximation (\ref{rhohighv}). 
\begin{figure}[t]
	\centering
	\includegraphics[width=\columnwidth]{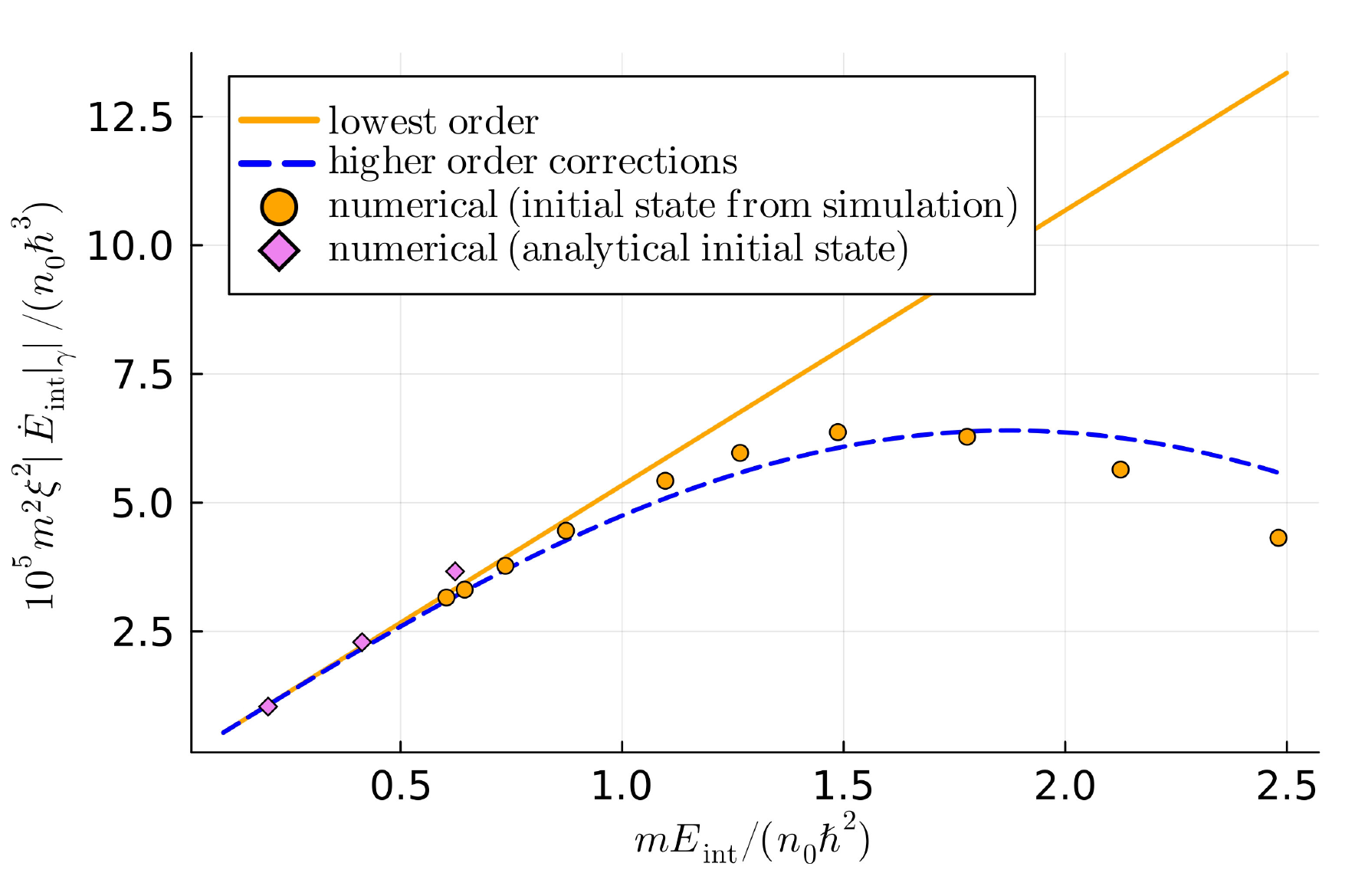}
	\caption{Damping rate of the interaction energy according to number damping. Shown are numerical results (orange dots and diamonds) as well as our analytical results (\ref{numberdamping}) (orange solid line) and (\ref{higherordernumber}) (blue dashed line). The damping rate features a minimum at $E_\text{int}\approx1.5n_0\hbar^2/m$.}
	\label{fig:deinthigherordernumber2}
\end{figure}
In this case, the spatial extent of the JRS is of the order of a few healing lengths $\xi$. This implies that the Fourier transform $\widetilde{n-n_0}$ will have an extent on the order $1/\xi$. Due to the derivatives, this is scaled by $\textbf{k}^2$ in the integral. 
Therefore, there is only a negligible contribution to the integral until $|\textbf{k}|\sim1/\xi$. We have $\tilde{\varepsilon}\propto F(|\textbf{k}l_z/2|^2)$, with $F$ again the scaled modified Bessel-function of the second kind. For sufficiently large $l_z$, it is thus allowed to use the asymptotical (large argument) expansion
\begin{align}
	\tilde{\varepsilon}(\textbf{k})&\sim 8a_\text{s}^2N_\text{cut}\frac{\sqrt{2\pi}}{|\textbf{k}|l_z}.
\end{align}
We hence can expect the damping rate to be antiproportional in $l_z$. Simulations for different values of $l_z$ imply that $l_z\sim\xi$ is already sufficiently large.

In section \ref{vortexannihilation} we discussed the decrease of interaction energy due to energy damping. We did not discuss number damping as we expect it to be recessive. The number damping rate takes a minimum at $E_\text{int}\approx1.5n_0\hbar^2/m$ (see figure \ref{fig:deinthigherordernumber2}). Afterwards, the curvature in interaction energy coincides with the higher order prediction found in \ref{higherordercorrections}. Hence, we do not expect an extremum in curvature.

\section{Dimensionless 2D SPGPE}
\label{Noise}
The full SPGPE contains not only the damping terms but also includes noise, which we neglected throughout. For each of the damping terms, one noise term appears. In this section we write down the dimensionless 2D SPGPE to clarify under which condition the neglect of noise in the main part can be expected to be valid.
	
	The noise term corresponding to energy damping is given by \cite{bradley_low-dimensional_2015}
\begin{align}
	i\hbar\textbf{(S)}d\psi|_{\varepsilon-\text{Noise}}=-\mathcal{P}\{\hbar\psi dU(\textbf{r},t)\},
\end{align}
where $\mathcal{P}$ is a projector on low energy modes, \textbf{(S)} signifies that the equation is in Stratonovich form and $dU$ is real noise with
\begin{align}
	\langle dU(\textbf{r},t)dU(\textbf{r}',t)\rangle=\frac{2k_\text{B}T}{\hbar}\varepsilon(\textbf{r}-\textbf{r}')dt.
\end{align}
In our units $\varepsilon$ has the dimensional prefactor $8a_\text{s}^2N_\text{cut}/\xi^2$ (see eq. (\ref{kernel})) and $dt$ gives us a dimensional factor $\hbar/\mu$.
	
The noise term corresponding to number damping is \cite{bradley_low-dimensional_2015}
\begin{align}
	i\hbar d\psi|_{\gamma-\text{Noise}}=i\hbar dW(\textbf{r},t),
\end{align}
where $dW$ is complex noise with
\begin{align}
	\langle dW(\textbf{r},t)dW(\textbf{r}',t)\rangle=\frac{2k_\text{B}T}{\hbar}\gamma\delta(\textbf{r},\textbf{r}')dt.
\end{align}
$\delta(\textbf{r},\textbf{r}')$ denotes the two-dimensional projected delta function with dimension $1/\xi^2$.
	
We obtain the full SPGPE by adding these terms and the projection on the coherent region $\mathcal{P}$ to equation (\ref{SPGPE}). By replacing $\textbf{r}\rightarrow\xi\textbf{r}$, $\textbf{k}\rightarrow\textbf{k}/\xi$, $t\rightarrow\hbar t/\mu$, $\psi\rightarrow\sqrt{n_0}\psi$, $dW\rightarrow\sqrt{2k_\text{B}T\gamma/\mu}dW$, $dU\rightarrow\sqrt{16a_\text{s}^2N_\text{cut}k_\text{B}T/\mu}dU$ we derive the dimensionless SPGPE
\begin{widetext}
	\begin{align}
	\begin{split}
		i(\textbf{S})d\psi&=\mathcal{P}\bigg\{[1-i\gamma]\left(-\Delta+(|\psi|^2-1)\right)\psi dt+\sqrt{\frac{8\pi}{n_0\lambda_\text{th}^2}\gamma}dW\\
		&-16n_0a_\text{s}^2N_\text{cut}\psi\int d^2\textbf{r}'\mathcal{F}^{-1}\left[F\left(\left|\frac{l_z\textbf{k}}{2\sqrt{2}\xi}\right|^2\right)\right](\textbf{r}-\textbf{r}')\nabla'\cdot\Im\{\psi^*\nabla'\psi\}dt-\psi\sqrt{\frac{8\pi}{n_0\lambda_\text{th}^2}8n_0a_\text{s}^2N_\text{cut}}dU\bigg\},\\
		&\langle dW^*(\textbf{r},t)dW(\textbf{r}',t)\rangle=\delta(\textbf{r},\textbf{r}')dt,\	\langle dU(\textbf{r},t)dU(\textbf{r}',t)\rangle=\mathcal{F}^{-1}\left[F\left(\left|\frac{l_z\textbf{k}}{2\xi}\right|^2\right)\right](\textbf{r}-\textbf{r}')dt.
	\end{split}
\end{align}
For completness we note again
\begin{align}
	\begin{split}
		&F(z)=e^zK_0(z),\ \mathcal{F}^{-1}\left[F\left(\left|\frac{l_z\textbf{k}}{2\xi}\right|^2\right)\right](\textbf{r}-\textbf{r}')=\int \frac{d^2\textbf{k}}{(2\pi)^2}e^{i\textbf{k}\cdot(\textbf{r}-\textbf{r}')}F\left(\left|\frac{l_z\textbf{k}}{2\xi}\right|^2\right),\\
		\gamma&=\frac{8n_0a_\text{s}^2}{n_0\lambda_\text{th}^2}e^{\beta\mu_{3\text{D}}}\int_0^1dy\ln\left(\frac{1-zy}{1-z}\right)\frac{1}{(1-y)(1-zy)},\ z=e^{\beta(\mu_{3\text{D}}-2\epsilon_\text{cut})},\ N_\text{cut}=\frac{1}{e^{\beta(\epsilon_\text{cut}-\mu_{3\text{D}})}-1}.
	\end{split}
\end{align}
\end{widetext}

As the noise terms are scaled by the inverse two-dimensional phase space density $1/(n_0\lambda_\text{th}^2)$ compared to their respective damping terms, the neglect of noise in the SPGPE is justified for a large phase space density\footnote{}. We remark, however, that according to section \ref{comparison} number damping is only dominant in the low phase space density regime. Hence, while for high phase space density a description with pure energy damping (without noise and number damping) is physical, if number damping is relevant, so is noise.

%


\end{document}